\documentclass[manuscript]{aastex}  
\usepackage{epstopdf}
\usepackage{ulem}
\usepackage{amssymb}
\usepackage{natbib}
\usepackage{times}
\usepackage{graphicx}
\usepackage[usenames,dvipsnames]{pstricks}
 \usepackage{psfrag}
                                                                                        
\def\apj{ApJ}%% Astrophysical Journal
\def\aap{A\&A}%% Astronomy and Astrophysics
\def\mnras{MNRAS}%% Monthly Notices of the RAS
\newcommand{\be}{\begin{equation}}
\newcommand{\ee}{\end{equation}}
\newcommand{\bse}{\begin{subequations}}
\newcommand{\ese}{\end{subequations}}
\newcommand{\bary}{\begin{eqnarray}}
\newcommand{\eary}{\end{eqnarray}}

\shorttitle{Neutrino Oscillations in Highly Magnetized }
\shortauthors{Fraija N.}

\begin{document}
\title{Propagation and neutrino oscillations in the base of a highly magnetized  gamma-ray burst fireball flow}
\author{N. Fraija\altaffilmark{1}}
\affil{Instituto de Astronom\'ia, UNAM, M\'exico, 04510}
\email{nifraija@astro.unam.mx}
\altaffiltext{1}{Luc Binette-Fundaci\'on UNAM Fellow. Instituto de Astronom\' ia, Universidad Nacional Aut\'onoma de M\'exico, Circuito Exterior, C.U., A. Postal 70-264, 04510 M\'exico D.F., M\'exico}

%\date{\today} 

\begin{abstract}
Neutrons play an important role in the dynamics of gamma-ray bursts.   The presence of neutrons in the baryon-loaded fireball is expected. If the neutrons abundance is comparable to that of  protons, important features may be observed such as  quasi-thermal multi-GeV neutrinos in coincidence with a subphotospheric $\gamma$-ray emission,   nucleosynthesis  at later times and rebrightening of the afterglow emission.  Additionally, thermal MeV neutrinos are created by electron-positron annihilation,  electron (positron) capture on protons (neutrons) and nucleonic bremsstrahlung. Although MeV neutrinos are difficult to detect, quasi-thermal GeV neutrinos are expected in km$^3$ detectors and/or DeepCore+IceCube.  In this paper, we show that neutrino oscillations have outstanding  implications for the dynamics of the fireball evolution and also that they can be detected through their flavor ratio on Earth.  For that we derive the resonance and charged-neutrality conditions as well as the neutrino self-energy and effective potential  up to order $m_W^{-4}$ at strong, moderate and weak magnetic field approximation to constrain the dynamics of the fireball.  We found important implications: i) resonant oscillations are suppressed for high baryon densities as well as neutrons abundance larger than that of protons and   ii)  the effect of magnetic field is to decrease the proton-to-neutron ratio aside from the number of multi-GeV neutrinos expected in DeepCore detector.   Also we estimate the GeV neutrino flavor ratios  along the jet  and  on Earth.
\end{abstract}


\section{Introduction}
Gamma-ray bursts (GRBs) are brief events occurring at an average rate of some days throughout the universe. These short, energetic bursts of gamma-rays mark the most violent, cataclysmic explosions in the universe, likely associated with the birth of stellar-size black-holes or rapidly spinning, highly magnetized neutron stars. One of  the most successful theory in terms of explaining GRBs and their afterglows is the fireball model \citep[see][for recent reviews]{mes06, zha04}.  This model predicts  an  expanding  ultrarelativistic shell that moves into the external surrounding medium.  The collision of the expanding shell  with another shell (internal shocks) or the interstellar medium (external shocks) gives rise to radiation emission through the leptonic (synchrotron and synchrotron self-Compton (SSC) radiation) or hadronic (photo-pion decay and inelastic proton-neutron collision) processes.  In addition,  when  the expanding relativistic shell encounters the external medium two shocks are involved: an outgoing, or forward shock \citep{ree94,pac93} and another one that propagates back into the ejecta, the reverse shock  \citep{mes94, mes97a}.\\
The base of the fireball flow is connected to the GRB central engine, a black hole (BH) - torus system or a rapidly rotating magnetar.  It is endowed with magnetic fields and formed by free nucleons, $e^{\pm}$ pairs in thermal equilibrium  of $1-10$ MeV inside the initial scale $\sim$10$^7$ cm assuming a 10$M_\odot$ BH. In the initial state, the baryons are essentially at rest with respect to the central engine \citep{zha04}. \\
In recent years, two observational facts have supported the idea that the central engine could be magnetized. i) The strong gamma-ray polarization \citep{cob03, bog03,rut04} and ii) a stronger magnetic field in reverse than in forward shock (magnetization of the jet) as a result of a full description of GRBs 980923, 990123, 021004, 021211, 041219A , 090926A and others \citep{zha02, zha03a, fan05,fra12a,fra12b,sac12}, so that the fireball may be endowed with a primordial magnetic field \citep{zha03a,zha05}.  Recently, a relevant result based on the magnetization of outflow and the non detection of neutrinos from GRBs by IceCube \citep{abb12b} was found by   \citet{zha13}.  They proposed that the suppression of the neutrino flux  in the gamma-ray emission region is a consequence of the degree of magnetization of the outflow.\\
The evolution of the fireball with  magnetic fields has been explored by various authors \citep{uso99,whe00,bla02,lyu02,spr03,vla03a,vla03b,zha11}. In these models an electromagnetic component  is introduced  in the dynamics of the standard fireball. This extra component is taken into account through the magnetization parameter ($\sigma$), defined as the ratio of Poynting flux (electromagnetic component) to matter energy (internal+kinetic component).  Depending on the dissipation process,  a part of the dissipated magnetic energy is converted into internal energy, and the other part is used to accelerate the fireball \citep{mic69, mes12}.\\
The energy requirement (isotropy-equivalent luminosities $L_\gamma \geq 10^{52}$ erg s$^{-1}$) demands magnetic fields at the base in excess of $B\sim 10^{15}$ G, that can be produced by shear and instabilities in an accreting torus around the BH. The energy source can be either the accretion energy or via magnetic coupling between the disk and BH, extraction of angular momentum from the latter occurring via the Blandford-Znajek mechanism \citep{bla77, mes97b, uso92, lee00,put01}. \\
On the other hand, the presence of neutrons and their abundance in the fireball GRB plays an important role  in the dynamics of the jet and its relation with the observed photon and neutrino flux emission \citep{koe07, met08,raz06a,der06}.  Among the more relevant consequences  are: i)  generation of quasi-thermal GeV neutrinos and subphotospheric $\gamma$-ray emission  via inelastic collisions between streaming protons and neutrons in the fireball. Above the pion production threshold ($\sim$140 MeV), the n-p decoupling is followed by inelastic n-p scattering leading to $\pi^{\pm}$ and $\pi^0$. As known, $\pi^0$ decays to two photons ($\gamma+\gamma$) and $\pi^+$ to four particles (three neutrinos) ($\mu^++\nu_\mu$ and then $\mu^+$ to $e^++\nu_e+\bar{\nu}_\mu$), which is the first signature in photons and neutrinos \citep{der99a,  bac00, mes00, raz06b, koh13}, ii) the effects on the development, strength and energy range of internal shocks \citep{ros06, xue08,fan05,fan04}, iii) nucleosyntesis \citep{met08,bel03a}, and iv) rebrightening  of the afterglow emission.  Neutrons decay after 881.5 s, hence they play an important role at large radii $\sim 10^{16} - 10^{17}$ cm where external shocks happen as even an exponential small number of survival neutrons carry an energy much larger than the rest-energy of circumburst medium \citep{der99b, bel03a, bel03b,ros06}.\\
Furthermore,  in the initial stage  and during the expansion of the fireball thermal neutrinos will be produced by electron-positron annihilation ($e^++e^-\to Z\to \nu_j+\bar{\nu}_j$),    processes of electron capture on protons   ($e^-+p  \to n+  \nu_e$) and positron capture on neutrons ($e^++n  \to p+  \bar{\nu_e}$), and  nucleon-nucleon bremsstrahlung ($NN\to NN+\nu_j+\bar{\nu}_j$) for $j=e,\nu,\tau$.    Also neutrinos of similar energies are naturally expected in the accretion disk during the collapse or merger.   
As known, properties of neutrinos get modified when they propagate in this magnetized fireball and,  depending on the neutrino flavor, would  feel a different effective potential because   electron neutrino ($\nu_e$) interacts with electrons via both neutral and charged currents (CC), whereas muon ($\nu_\mu$) and tau ($\nu_\tau)$ neutrinos interact only via the neutral current (NC). This would induce a coherent effect in which maximal conversion of $\nu_e$ into $\nu_\mu$ ($\nu_\tau$) takes place even for a small intrinsic mixing angle \citep{wol78}.   Although  thermal neutrino oscillations  have been widely studied in the literature for different scenarios \citep{ruf99,goo86,vol99,das08,jan12,sah05, dol96b,erd00, dol97,dol03, erd09, sah09a, sah09b}, and even for O(m$^{-4}_W$) corrections to the neutrino dispersion relation \citep{not88,enq91,dol92},  a full analysis of both the thermal and subphotospheric neutrino oscillations over the dynamics of the fireball GRB as neutrons abundance is comparable to that of  protons and the magnetic field is as strong as the critical magnetic field  has not been performed.     In this paper, we show that neutrino oscillations have important implications for the evolution of the fireball when it is highly magnetized and has a considerable amount of neutrons and protons.   Firstly, we compute the neutrino self-energy  by using real time formalism of finite temperature field theory \citep{nie90,wel82,dol03,erd90} and Schwinger's proper-time method \citep{sch51}. Then, taking into account the strong, moderate and weak magnetic field approximation, we compute, analyze and compare the neutrino effective potential up to order m$^{-4}_W$ at all the regimes.  Next, we derive the resonance conditions to find the range of values  for which MeV - GeV neutrinos oscillate resonantly and then we do a full analysis of two- (solar, atmospheric and accelerator parameters)  and three-neutrino mixing, even though we estimate the neutrino flavor ratio in the jet and on Earth.  After that, we derive the charged-neutrality condition so that together with the resonance condition we can constrain the baryon density as well as  the  protons and neutrons abundance in the dynamics and  evolution of a magnetized fireball. The results are also discussed in terms of magnetization degree.\\
We hereafter use $Q_x\equiv Q/10^x$ in cgs units and $c=\hbar$=k=1 in natural units 
\section{Neutrino Effective Potential}
We use the finite temperature field theory formalism to study the effect of heat bath  on the propagation of elementary particles \citep{dol96a,tut02}. The effect of magnetic field is taken into account through Schwinger's propertime method \citep{sch51}. The effective potential of a particle is calculated from the real part of its self-energy diagram. The neutrino field equation  in a magnetized medium is
\be
[ {\rlap /k} -\Sigma(k) ] \Psi_L=0,
\label{disneu}
\ee
where the neutrino self-energy operator $\Sigma(k)$ is a Lorentz scalar which depends on the characterized parameters of  the medium, for instance,  chemical potential, particle density, temperature, magnetic field, etc.  For our purpose, $\Sigma(k)$ can be formed by
\be
\Sigma (k) = {\mathcal R} \Big( a_\parallel {\rlap /k}_\parallel
+ a_\perp {\rlap /k}_\perp + b {\rlap /u} + c {\rlap /b} \Big){\mathcal L} \,,
\label{sigmanb}
\ee
where $k^\mu_\parallel=(k^0, k^3)$, $k^\mu_\perp=(k^1,k^2)$ and $u^\mu$ stands for the 4-velocity of the center-of-mass of the medium given by $u^\mu = (1, {\bf 0})$.  The projection operators are conventionally defined as $ {\mathcal R}=\frac12(1+\gamma_5)$ and $ {\mathcal L} = \frac12(1-\gamma_5)$.
The effect of the magnetic field enters through the 4-vector $b^\mu$ which is given by $b^\mu = (0, {\hat {\bf b}})$. The background classical magnetic field vector is along
the $z$-axis and consequently $b^\mu=(0,0,0,1)$. So using the four vectors  $u^\mu$ and  $b^\mu$ we can express
\be
{\rlap /k}_\parallel= k_0 {\rlap /u} - k_3 {\rlap /b},
\label{paralk}
\ee
and the self-energy can be expressed in terms of three independent four-vectors $k^\mu_\perp$, $u^{\mu}$ and $b^{\mu}$. Therefore we can write ($\Sigma =  {\mathcal R} {\tilde\Sigma}  {\mathcal L}$)
\be
{\tilde \Sigma}=a_{\perp}\rlap /k_{\perp}+b\rlap /u+c\rlap /b.
\label{sigmanb2}
\ee
The neutrino self-energy in a magnetic background can be found from Eq. (\ref{disneu}) 
\be
det[ {\rlap /k} -{\Sigma}(k)]=0.
\ee
Using the Dirac algebra, the dispersion relation, $V_{eff}=k_0-|{\bf k}|$, as a function of Lorentz scalars can be written as 
\be
V_{eff}=b-c\,\cos\phi-a_{\perp}|{\bf k}|\sin^2\phi,
\label{poteff}
\ee
where $\phi$ is the angle between the neutrino momentum and the magnetic field vector.   Now the Lorentz scalars $a$, $b$ and $c$ which are functions of neutrino
energy, momentum and magnetic field can be calculated from the neutrino self-energy due to CC and NC interactions of neutrino with the background particles.
\subsection{One-loop  neutrino self-energy}
Let us consider one-loop corrections to the neutrino self-energy in the presence of a magnetic field.  The one-loop neutrino self-energy  comes from three pieces  \citep{bra07, eli04,erd98,sah09a, sah09b}, one coming from the $W$-exchange diagram which we will call $\Sigma_W (k)$ (Fig. 1(a)), one from the $Z$-exchange diagram which will be denoted by $\Sigma_Z (k)$ (Fig. 1(b)) and one from the tadpole diagram which we will designate by $\Sigma_t (k)$ (Fig. 1(c)). The total neutrino self-energy  in a magnetized medium then becomes
\be
\Sigma(k) = \Sigma_W(k) + \Sigma_Z (k)+ \Sigma_t (k)\,.
\label{tsen}
\ee
The W-exchange diagram for the one-loop self-energy is 
\be
-i\Sigma_W(k)= {\mathcal R}\,\biggl[\int\frac{d^4 p}{(2\pi)^4}\left(\frac{-ig}{\sqrt{2}}\right)
\gamma_\mu\,  \,iS_{\ell}(p)\left(\frac{-ig}{\sqrt{2}}\right)\gamma_\nu
 \,i W^{\mu \nu}(q)\biggr]\, {\mathcal L}\,,
\label{Wexch}
\ee
where $g^2=4\sqrt2 G_Fm_W^2$  is the weak coupling constant, $W^{\mu\nu}$ depicts the W-boson propagator which  in the eB $\ll$ m$^2_W$ limit and in unitary gauge is given by \citep{erd98,sah09b}
\be
W^{\mu\nu}(q)=\frac{g^{\mu\nu}}{m^2_W}\biggl(1+\frac{q^2}{m^2_W}  \biggr)-\frac{q^\mu q^\nu}{M^4_W}+\frac{3ie}{2m^4_W}F^{\mu\nu}\,,
\ee
here m$_W$ is the W-boson mass, $g^{\mu\nu}$ is the  metric tensor and $F^{\mu\nu}$ is the electromagnetic field tensor. S$_l$(p) stands for the charged lepton propagator which can  be separated in to two  charged propagators; one in presence of a uniform background magnetic field ($S^0_\ell(p)$) and the other in a magnetized medium ($S^\beta_\ell (p)$) and can be written as 
\be
S_\ell (p) = S^0_\ell(p) + S^\beta_\ell (p)\,.
\label{slp}
\ee
Assuming that the z-axis points in the direction of the magnetic field ${\rm B}$, we can express the charged lepton propagator  in presence of a uniform background magnetic field  as
\be
i S^0_\ell(p) = \int_0^\infty e^{\phi(p,s)} G(p,s)\,ds\,,
\label{sl0p}
\ee
where  the functions $\phi(p,s)$ and $G(p,s)$ are give by
\bary
\phi(p,s) &= &is(p_0^2 - m_\ell^2) - is[p^2_3 + \frac{\tan z}{z} p^2_\perp]\,,\cr
G(p,s)&=&\sec^2 z \left[{\rlap A/} + i {\rlap B/} \gamma_5 +m_\ell(\cos^2 z -
i \Sigma^3 \sin z \cos z)\right]\,,
\label{phaselu}
\eary
where $m_l$ is the mass of the charged lepton, $p^2_\parallel = p_0^2 - p_3^2$ and  $p^2_\perp = p_1^2 + p_2^2$ are the projections  of the momentum on the magnetic field direction  and  $z= e{ B}s$, being $e$ the magnitude of the electron charge. Additionally,  the covariant  vectors are given as follows $A_\mu = p_\mu -\sin^2 z (p\cdot u\,\, u_\mu - p\cdot b \,\,b_\mu)\,$, $B_\mu = \sin z\cos z (p\cdot u \,\,b_\mu - p\cdot b \,\,u_\mu)\,,$ and $\Sigma^3 = \gamma_5 {\rlap /b} {\rlap /u}\,.$\\ 
On the other hand, the charged lepton propagator in a magnetized medium is given by
\be
S^\beta_\ell(p) = i \eta_F(p\cdot u)\int_{-\infty}^\infty e^{\phi(p,s)} G(p,s)
\,ds\,,
\label{slbp}
\ee
here $\eta_F(p\cdot u)$ contains the distribution functions of the particles in the medium which can be written as
\be
\eta_F (p \cdot u) = \frac{\theta(p\cdot u)}{e^{\beta(p\cdot u -
\mu_\ell)} + 1 } +
\frac{\theta(- p\cdot u)}{e^{-\beta(p\cdot u - \mu_\ell)} + 1}\,,
\label{eta}
\ee
where $\beta$ and $\mu_\ell$ are the inverse of the medium temperature and the chemical potential of the charged lepton, respectively.\\
The Z-exchange diagram for the one-loop self-energy is 

\be
-i\Sigma_Z(k)= {\mathcal R}\,\biggl[\int\frac{d^4 p}{(2\pi)^4}\left(\frac{-ig} {\sqrt{2}\cos\theta_W}\right) 
\gamma_\mu \,iS_{\nu_\ell}(p)\left(\frac{-ig}{\sqrt{2}\cos\theta_W}\right)
\gamma_\nu\,i Z^{\mu \nu}(q)\biggr]\, {\mathcal L}\,,
\label{Zexch}
\ee
where $\theta_W$ is the Weinberg angle, Z$^{\mu\nu}$(q) is the Z-boson propagator in vacuum, S$_{\nu_l}$ is the neutrino propagator in a thermal bath of neutrinos. \\
The Tadpole diagram for the one-loop self-energy is 

\be
i\Sigma_t(k)= {\mathcal R}\,\biggl[ \left(\frac{g}{2\cos \theta_W}\right)^2\,
\gamma_\mu\,iZ^{\mu \nu}(0)\int\frac{d^4 p}{(2\pi)^4} {\rm Tr}
\left[\gamma_\nu \,(C_V + C_A \gamma_5)\,iS_{\ell}(p)\right]\,\biggr] {\mathcal L}\,,
\label{tad}
\ee
where the quantities $C_V$ and $C_A$ are the vector and axial-vector coupling constants which come in the neutral-current interaction of
electrons, protons ($p$), neutrons ($n$) and neutrinos with the $Z$ boson.  Their forms are as follows
\bary
C_V=\left \{\begin{array}
{r@{\quad\quad}l}
-\frac{1}{2}+2\sin^2\theta_W & e\\
\frac{1}{2} & {\nu}\\ \frac{1}{2}-2\sin^2\theta_W & {{p}}\\
-\frac{1}{2} & {{n}}
\end{array}\right.,
\label{cv}
\eary
and
\bary
C_A=\left \{\begin{array}
{r@{\quad\quad}l}
-\frac{1}{2} & {\nu},{{p}}\\
\frac{1}{2} & e,{{n}}
\end{array}\right..
\label{ca}
\eary
By evaluating eq. (\ref{Wexch}) explicitly we obtain
\be
Re \Sigma_W(k)= {\mathcal R}\,[a_{W_\perp} \rlap /k_\perp + b_W \rlap /u + c_W \rlap /b]\, {\mathcal L}\,,
\ee
where the Lorentz scalars are given by
\bary
a_{W\perp}&=&-\frac{\sqrt2G_F}{m_W^2}\biggl[
\biggl\{E_{\nu_e}(n_e-\bar{n}_e)+ k_3(n_e^0-\bar{n}_e^0)\biggr\}\nonumber\\
&&+\frac{eB}{2\pi^2}\int^\infty_0 dp_3\sum_{n=0}^\infty(2-\delta_{n,0})
\biggl (\frac{m_e^2}{E_n}- \frac{H}{E_n}\biggr)(f_{e,n}+\bar{f}_{e,n})\biggr],
\label{conaw}
\eary
\bary
b_W&=& \sqrt2 G_F \biggl[\biggl(1+\frac32\frac{m_e^2}{m_W^2}
+\frac{E_{\nu_e}^2}{m_W^2}\biggr)(n_e-\bar{n}_e)+\biggl(\frac{eB}{m_W^2}
+\frac{ E_{\nu_e}k_3}{m_W^2}\biggr)(n_e^0-\bar{n}_e^0)\nonumber\\
&&-\frac{eB}{2\pi^2m_W^2} \int^\infty_0 dp_3
\sum_{n=0}^\infty(2-\delta_{n,0})\biggl\{2k_3E_n\delta_{n,0}
+2E_{\nu_e}\biggl(E_n-\frac{m_e^2}{2E_n}\biggr)\biggr\}(f_{e,n}+\bar{f}_{e,n})\biggr]
\label{conbw}
\eary
and
\bary
c_W&=&\sqrt2
G_F\biggl[\biggl(1+\frac12\frac{m_e^2}{m_W^2}-\frac{k_3^2}{m_W^2}\biggr)(n_e^0-\bar{n}_e^0)
+\biggl(\frac{eB}{m_W^2}-\frac{E_{\nu_e}k_3}{m_W^2}\biggr)(n_e-\bar{n}_e)\nonumber\\
&&-\frac{eB}{2\pi^2m_W^2} \int^\infty_0
dp_3\sum_{n=0}^\infty(2-\delta_{n,0})\biggl\{2E_{\nu_e}
\biggl(E_n-\frac{m_e^2}{2E_n}\biggr)\delta_{n,0}\cr
&&+2k_3\biggl(E_n-\frac32\frac{m_e^2}{E_n}-
\frac{H}{E_n}\biggr)\biggr\}(f_{e,n}+\bar{f}_{e,n})\biggr].
\label{concw}
\eary
where the electron number density  and electron distribution function are
\be
n_e(\mu, T, B)=\frac{eB}{2\pi^2}\sum_{n=0}^\infty (2 - \delta_{n,0}) \int_0^\infty  \frac{dp_3}{e^{\beta(E_{e,n}-\mu)} +1},
\label{den}
\ee
and
\be
f(E_{e,n},\mu)=\frac{1}{e^{\beta(E_{e,n}-\mu)} +1}\,,
\label{den1}
\ee
respectively, with $\bar{f}_{e,n}(\mu, T)=f_{e,n}(-\mu, T)$ and  $E_{e,n}=\sqrt{p_3^2+m_e^2+H}$ with $H=2neB$. We can also express the Eq.~(\ref{Zexch}) (Z-exchange) as
\be
Re\Sigma_Z(k) = {\mathcal R}(a_Z\rlap /k+b_Z \rlap /u) {\mathcal L}\,,
\label{sigz}
\ee
and explicit evaluation gives \citep{dol94}
\be
a_Z=\sqrt{2}G_F\biggl[\frac{E_{\nu_e}}{m_Z^2}(n_{\nu_e}-\bar{n}_{\nu_e})
+ \frac23\frac{1}{m_Z^2}\biggl(\langle E_{\nu_e}\rangle n_{\nu_e}
+\langle \bar{E}_{\nu_e}\rangle \bar{n}_{\nu_e}\biggr)\biggr],
\ee
and
\be
b_Z=\sqrt{2}G_F\biggl[(n_{\nu_e}-\bar{n}_{\nu_e})-\frac{8E_\nu}{3m^2_Z}
\biggl(\langle E_{\nu_e}\rangle n_{\nu_e}
+\langle \bar{E}_{\nu_e}\rangle\bar{n}_{\nu_e}\biggr)\biggr],
\ee
where the four vector $\rlap /k$ can be decomposed  into the four vectors $\rlap /u$ and $\rlap /b$  in accordance with Eq.~(\ref{paralk}).\\
From the tadpole diagram, Eq.~(\ref{tad}), we obtain
\bary
Re\Sigma_t (k) &=&\sqrt2 G_F {\mathcal R}\biggl[
\biggl\{C_{V_e}(n_e-\bar{n}_e)+C_{V_p}(n_p-\bar{n}_p)+C_{V_n}(n_n-\bar{n}_n)
+(n_{\nu_e}-\bar{n}_{\nu_e})
\nonumber\\
&&+(n_{\nu_\mu}-\bar{n}_{\nu_\mu})+(n_{\nu_\tau}-\bar{n}_{\nu_\tau})\biggr\} \rlap/u
-C_{A_e} (n_e^0-\bar{n}_e^0)\rlap /b\biggr] {\mathcal L}.
\eary
For anti-neutrinos we must change (n$_x$-$\bar{n}_x$) by  -(n$_x$-$\bar{n}_x$).    The different contributions to the neutrino self-energy up to order the $1/m^4_W$ have been calculated in a background of $\gamma$, $e^\pm$, free baryons, neutrinos and anti-neutrinos. The effective potential that is applicable to neutrino oscillations in matter is $V_{eff}=V_e-V_{\mu,\tau}$ which depends only on electron density \citep{wol78, dol92}, assuming  that  neutrinos propagate in the same direction of the magnetic field ($\phi=0$) and with k$_3$= E$_{\nu_e}$,  we derive the neutrino effective potential in all  the regimes of the magnetic field:  strong  $\Omega_B= eB/m_e^2\gg 1$, moderate $\Omega_B=eB/m_e^2 > 1$ (above $B_c$) and $\Omega_B=eB/m_e^2 \leq 1$ (below $B_c$), and weak  $\Omega_B=eB/m_e^2 \ll 1$ regime, where $B_c=m_e^2/e=4.414\times 10^{13}$ G is the critical magnetic field.   In the following subsections we will show the neutrino effective potential in all the regimes (strong, moderate and weak magnetic field limit)\citep{erd09}; the calculations are explicitly shown in the appendix. Using the typical fireball values at the initial stage and the phase of acceleration;  temperature in the range 1- 10 MeV at initial radius r$_0\simeq 10^{6.5} -10^{7.5}$ cm  for thermal MeV neutrinos \citep{bel03a,ruf99,koe05,jan12} and in the range 0.1 - 1 MeV  at radius r$\simeq 10^{8.5} -10^{9.5}$ cm for quasi-thermal GeV neutrinos  \citep{bac00, mes00,raz06a,raz06b,koh13},  we plot the effective potential in each limit as shown in figs. \ref{Bstrong}, \ref{Babove}, \ref{Bbelow} and \ref{Bweak}. Although we are going to do a full analysis of neutrino oscillations in a fireball endowed with moderate magnetic fields which is more favored for GRB central engines \citep{bel03a,mes12},  a brief and additional description  at  the strong and weak field limit as well as a comparison of effective potentials at all the regime  will be given as follows.\\
\subsection{Strong Magnetic field: $\Omega_B \gg 1$}
In the strong magnetic field approximation ($m^2_e \ll T^2\ll \Omega_B\,m^2_e$), leptons are all confined  to the lowest Landau level (n=0), then only this level will contribute to the potential and the energy of these leptons will be independent of the magnetic field.  The neutrino effective potential at the strong magnetic field is given by
\bary
V_{eff,S}=A_e \biggl[\sum^\infty_{l=0} (-1)^l \sinh\alpha_l\, K_1(\sigma_l)\biggl\{\frac{m_e^2}{m^2_W}\biggl(1+4\frac{E^2_\nu}{m^2_e}\biggr)\biggr\} -3\frac{m_e^2}{m^2_W}\frac{E_\nu}{m_e}\sum^\infty_{l=0} (-1)^l \cosh\alpha_l K_0(\sigma_l)\biggr].\cr
\label{Spotef}
\eary
In fig. \ref{Bstrong},  we plot the effective potential at the strong field limit (eq. $\ref{Spotef}$) as a function of temperature (left-hand figure above), magnetic field (right-hand figure above) and chemical potential (figure below). For these plots we take into account the neutrino energy 10 MeV and  the values of temperature, magnetic field and chemical potential in the  range 1 to 10 MeV,  $10^3$ to $10^5\,B_c$ and 10$^{-4}$ eV to 4.5 keV, respectively.  As shown, the effective potential is a quasi-constant function of temperature, and an increasing function of magnetic field and chemical potential 
\subsection{Moderate Magnetic field: $\Omega_B > 1$ and $\Omega_B \leq 1$  }
In the moderate field approximation ($m^2_e < \Omega_B\,m^2_e\leq T^2$),  leptons start to occupy the next Landau levels (n=1, 2, 3 ..) which have a separation  that is directly proportional to magnetic field.  For this case each of these levels will contribute to the effective potential  and the energy  of leptons are directly proportional to magnetic field.  The neutrino effective potential in the moderate magnetic field is written as
\bary\label{MApotef}
V_{eff}=A_e\biggl[\sum^\infty_{l=0} (-1)^l \sinh\alpha_l\,\biggl\{\frac{m_e^2}{m^2_W}\biggl(1+4\frac{E^2_\nu}{m^2_e}\biggr)K_1(\sigma_l)+\sum^{\infty}_{n=1}\lambda_n\,\biggl( 2+\frac{m_e^2}{m^2_W}\biggl( 3-2\Omega_B+4\frac{E^2_\nu}{m_e^2}\biggr)\biggr)\,K_1(\sigma_l\lambda_n) \biggr\}\cr\nonumber
-4\frac{m_e^2}{m^2_W}\frac{E_\nu}{m_e}\sum^\infty_{l=0} (-1)^l \cosh\alpha_l\biggl\{ \frac34K_0(\sigma_l)+\sum^{\infty}_{n=1}\lambda^2_n\, K_0(\sigma_l\lambda_n)  \biggr\}\biggr]\,,\hspace{5.cm}
\eary
with
\begin{equation}\label{MApotef}
\lambda^2=
\cases{
2\,n\,\Omega_B& for  $\,\Omega_B>$ 1\cr
1+2\,n\,\Omega_B & for    $\,\Omega_B\leq$ 1\..\cr
}
\end{equation}
In fig. \ref{Babove}, we have plotted the neutrino effective potential at the moderate field limit above $B_c$ (eq. \ref{MApotef}) as a function of temperature (left-hand figure above),  magnetic field (right-hand figure above) and  chemical potential (figure below).  For these plots we take into account the neutrino energy 10 MeV and  the values of temperature, magnetic field and chemical potential in the  range 1 to 10 MeV,  $B_c < B \leq 10^2\,B_c$ and 10$^{-4}$ eV  to 4.5 keV, respectively.  As shown, the effective potential as a function of temperature has two  different behaviors. Firstly, it  is an increasing function of temperature in the range 1 to 3 MeV for values of magnetic field $5 B_c$ and $10 B_c$, and in the range 1 to 7 MeV for  $50 B_c$ and $100 B_c$ and secondly, it tends to be constant for values of  temperature greater than  3 MeV for $B=5 B_c$ and $B=10 B_c$, and 7 MeV for  $B=50 B_c$ and $B=100 B_c$. In this figure  it is also seen that the effective potential is an increasing function for both the magnetic field and chemical potential. \\
In fig. \ref{Bbelow}, we have plotted the neutrino effective potential at the moderate field limit below B$_c$ (eq. \ref{MApotef}) as a function of temperature (figure above),  magnetic field (middle figure) and  chemical potential (figure below) for neutrino energies $E_\nu=10$ (left column figures)   MeV and  $E_\nu=10$ GeV (left column figures). For these plots we take into account  the values of  the magnetic field and chemical potential in the  range  $10^{-3} B_c < B \leq \,B_c$ and 10$^{-4}$ eV to 4.5 keV, respectively, and two ranges of temperatures;   0.1 MeV $\leq T\leq$ 1 MeV (right column figures) and 1 MeV $\leq T\leq$ 10 MeV (left column figures). In the top figures, the behavior of the effective potential as a function of temperature has a multifunctional dependence  which depends on  the values of the magnetic field. For instance, in the right-hand figure,  for B=0.5 B$_c$ it is a dramatically increasing function,   for B=10$^{-2}\,B_c$  it is just a steadily increasing function and for the smaller values of  B, it becomes a decreasing function and in the left-hand figure,  for B=$0.5\,B_c$ it is a steadily increasing function and as B decreases  the effective potential gradually becomes a decreasing function. In the middle figures, the effective potential is a dramatically (left-hand figure) and (right-hand figure) steadily increasing function of the magnetic field regardless of the values of temperature. In the bottom figures, the effective potential represented by means of  an increasing function of the chemical potential shows the same behavior but at different energy ranges. \\
\subsection{Weak Magnetic field: $\Omega_B \ll 1$}
In the weak field approximation ($ \Omega_B\,m^2_e\ll m^2_e $),  all  Landau levels are occupied  and  overlapped each other, therefore  a good description of these levels is to take the approximation $\sum_n\to \int dn$.  For this case, the neutrino effective potential is
\bary
V_{eff,W}= A_{e}\,\biggl[\sum^\infty_{l=0} (-1)^l \sinh\alpha_l\, \biggl\{ \biggl(   2+\frac{m_e^2}{m^2_W}\biggl(3+4\frac{E^2_\nu}{m^2_e}\biggr) \biggr) \biggl(\frac{K_0(\sigma_l)}{\sigma_l}+ 2\frac{K_1(\sigma_l)}{\sigma^2_l}  \biggr)\Omega^{-1}_B  -2\biggl(1+\frac{m^2_e}{m^2_W} \biggr) K_1(\sigma_l) \biggr\} \cr
-4\frac{m_e^2}{m^2_W}\frac{E_\nu}{m_e}\sum^\infty_{l=0} (-1)^l \cosh\alpha_l \biggl\{  \biggl( \frac{2}{\sigma_l^2\,\Omega_B}-\frac14\biggr) K_0(\sigma_l) +\biggl( 1+\frac{4}{\sigma^2_l} \biggr) \frac{K_1(\sigma_l)}{\sigma_l} \Omega_B^{-1}\biggl\}  \biggr]. \hspace{2.6cm}\cr
\label{Wpotef}
\eary
As shown in fig. \ref{Bweak} the neutrino effective potential  at the weak field limit  (eq. \ref{Wpotef})  as a function of temperature (figures above) and  chemical potential (figures below) for neutrino energies E$_\nu$=10 MeV (left-hand figures) and E$_\nu$=10 GeV (left-hand figures) is plotted. Due to the strength of the magnetic field it is quite small and any variation will produce insignificant changes in the effective potential; we only show the magnetic field contribution, i.e. by subtracting the effective potential with B=0, which shows that the potential is  a decreasing function of temperature and  an increasing function of chemical potential, and that  the magnetic contribution is the opposite as compared with the medium contribution. The effective potential in this regime differs from that  calculated by \citet{sah09b}, because the authors  took the solution of dispersion relation $k_3=-E_{\nu_e}$ instead of  $k_3=E_{\nu_e}$.
\subsection{Comparison of effective potentials at all the regimes}
We plot the neutrino effective potentials as a function of temperature (left-hand figure) and chemical potential (right-hand figure) in a strong, moderate and weak magnetic field approximation (fig. 6) which are given in eqs. (\ref{Spotef}), (\ref{MApotef}) and (\ref{Wpotef}), respectively.   As shown in fig. \ref{Bcompar},  the effective potential lies between $\sim10^{-11}$  and $\sim 10^{-8}$ eV for a  temperature in the range 1 to  8 MeV. One can see from this plot several  features. i) at T$\sim$1 MeV, the effective potentials at moderate (above and below $B_c$) field limit are quite close at $\sim 5\times 10^{-10}$ eV, however as temperature increases the separation between them also increases and for strong and weak field limit, the distance between the effective potentials is more than two orders of magnitude.  ii) As temperature increases the effective potentials at the strong and moderate (above $B_c$) field limits become closer to each other.   Dividing this plot  in two regions,   T $<$ 3 MeV and  T$\geq$ 3 MeV,  we can argue that for the given values of the magnetic field and temperature,  in the first region   m$^2_e \Omega_B \simeq T^2$, even though being slightly lower at T$\sim$1.4 MeV and as temperature becomes larger than the magnetic field, m$^2_e \Omega_B \ll T^2$, a difference between both effective potentials for each value of temperature  is observed. This difference, which is  of the same order of magnitude,  comes from the  contribution  of excited  Landau levels (n=1,2 ..).    iii) The effective potential at the weak field limit is a dramatically increasing function of temperature  which tends at the moderate limit below B$_c$.  iv) The effective potential at strong field limit decreases very gradually (quasi constantly) as temperature increases, in fact for the range of temperature considered, the effective potential is almost invariant to any thermal contribution.   Also it is an increasing function of the chemical potential in the range 0.1- 4$\times 10^3$ eV for all the regimes of the magnetic fields. By comparison the effective potential  at  strong (maximum value) and weak (minimum value)  field approximation, one can see that the last approximation is smaller by three orders of magnitude than the first one.\\
Additionally, we plot the contribution of m$^{-4}_W$ terms to the neutrino effective potential as shown in fig. \ref{order4}. As can be seen in eqs. (\ref{Spotef}), (\ref{MApotef}) and (\ref{Wpotef}), there are two common terms which depend on m$^{-4}_W$ and that contribute to  the effective potentials ; $(-1)^l \cosh(\alpha) K_0(\sigma)$ and 4 $E_\nu^2/m_W^2$. In the first case,  we compare the terms  $(-1)^l \sinh(\alpha) K_1(\sigma)$ and $(-1)^l \cosh(\alpha) K_0(\sigma)$ (left-hand figure) where the first term comes from particle-antiparticle asymmetry ($n_e - \bar{n}_e$) and the second one from ($n_e + \bar{n}_e$). From this figure, one can observe that the $(n_e - \bar{n}_e)/(n_e + \bar{n}_e)$ ratio is directly proportional to chemical potential and for $\mu\leq 10^2$ eV the  m$^{-4}_W$ term begins to be dominant  achieving a minimum value of 7 orders of magnitude for $\mu=10^{-3}$ eV.  In other words, as $\mu$ decreases the correction of order m$^{-4}_W$ is important and dominant over m$^{-2}_W$ term, even though one can deduce that for $\mu=0$, $n_e - \bar{n}_e=0$, then the only contribution would come from the term O(m$^{-4}_W$).  In the second case we plot 4 $E_\nu^2/m_W^2$ as a function of E$_\nu$ (right-hand figure). From this plot, you can notice that this term in comparison with unity  starts to contribute for neutrinos with energies of some tens of GeV.  As shown above, these two terms with the values of temperature, chemical potential and neutrino energy  are relevant to the neutrino effective potential.\\
From the previous analysis  can be observed that regardless of the magnetic field limit or relativistic  temperature or chemical potential, the neutrino effective potential is positive, therefore  due to its positivity  ($V_{eff,k}>$ 0 for k=S, M and W),  MeV - GeV neutrinos can oscillate resonantly.\\
\section{Neutrino Mixing and  Resonance Condition}
In the following subsections we are going to consider the best fit values of the neutrino oscillation parameters for two-neutrino mixing (solar, atmospheric and accelerator) and three-neutrino mixing.
\subsection{Two-Neutrino Mixing}
Here we consider the neutrino oscillation process $\nu_e\leftrightarrow \nu_{\mu, \tau}$. The evolution equation for the propagation of neutrinos in the  medium is given by \citep{fra14a}
\be
i
{\pmatrix {\dot{\nu}_{e} \cr \dot{\nu}_{\mu}\cr}}
={\pmatrix
{V_{eff}-\Delta \cos 2\theta & \frac{\Delta}{2}\sin 2\theta \cr
\frac{\Delta}{2}\sin 2\theta  & 0\cr}}
{\pmatrix
{\nu_{e} \cr \nu_{\mu}\cr}},
\ee
where $\Delta=\delta m^2/2E_{\nu}$,   $\delta m^2$ is the mass difference, $V_{eff}$ is the neutrino effective potential between $V_{\nu_e}$ and $V_{\nu_{\mu, \tau}}$  (eq. \ref{MApotef}),    $E_{\nu}$ is the neutrino energy and $\theta$ is the neutrino mixing angle. The conversion probability for the above process at a time $t$ is given by
\be
P_{\nu_e\rightarrow {\nu_{\mu}{(\nu_\tau)}}}(t) = 
\frac{\Delta^2 \sin^2 2\theta}{\omega^2}\sin^2\left (\frac{\omega t}{2}\right
),
\label{prob}
\ee
with
\be
\omega=\sqrt{(V_{eff}-\Delta \cos 2\theta)^2+\Delta^2 \sin^2
    2\theta}.
\ee
The oscillation length for the neutrino is given by
\be
l_{osc}=\frac{l_v}{\sqrt{\cos^2 2\theta (1-\frac{V_{eff}}{\Delta \cos 2\theta})^2+\sin^2 2\theta}}\,,
\label{osclength}
\ee
where $l_v=2\pi/\Delta$ is the vacuum oscillation length.  Applying the resonance condition given by
\bary
V_{eff}&=&\Delta \cos 2\theta\cr
&=&5\times 10^{-7}\, eV\,\frac{\delta m^2_{eV}}{E_{\nu,MeV}}\,\cos2\theta\,,
\label{reso2}
\eary
we obtain that the resonance length ($l_{res}$) can be written as
\be
l_{res}=\frac{l_v}{\sin 2\theta}.
\ee
In eq. (\ref{reso2}) the neutrino effective potential  depends on the chemical potential, temperature, the neutrino energy and the oscillation parameters  (mass differences and the mixing angles).  We will use the following parameters for this analysis:\\
\textbf{Solar Neutrinos}: A two-flavor neutrino oscillation analysis yielded $\delta m^2=(5.6^{+1.9}_{-1.4})\times 10^{-5}\,{\rm eV^2}$ and $\tan^2\theta=0.427^{+0.033}_{-0.029}$\citep{aha11}.\\
\textbf{Atmospheric Neutrinos}: Under a two-flavor disappearance model with separate mixing parameters between neutrinos and antineutrinos the following parameters for the SK-I + II + III data $\delta m^2=(2.1^{+0.9}_{-0.4})\times 10^{-3}\,{\rm eV^2}$ and $\sin^22\theta=1.0^{+0.00}_{-0.07}$ were found.\citep{abe11a}.\\
\textbf{Accelerator Parameters (Short baselines)}: \cite{chu02} found two well defined regions of oscillation parameters with either $\delta m^2  \approx  7\, {\rm eV^2}$ or $\delta m^2 < 1\, {\rm eV^2} $ compatible with both LAND and KARMEN experiments, for the complementary confidence. In addition, MiniBooNE found evidence of oscillations in the 0.1 to 1.0 eV$^2$, which are consistent with LSND results \citep{ath96, ath98}.
\subsection{Three-Neutrino Mixing}
To determine the neutrino oscillation probabilities we have to solve the evolution equation of the neutrino system in the matter. In a three-flavor framework, this equation is given by
\be
i\frac{d\vec{\nu}}{dt}=H\vec{\nu},
\ee
and the state vector in the flavor basis is defined as
\be
\vec{\nu}\equiv(\nu_e,\nu_\mu,\nu_\tau)^T.
\ee
The effective Hamiltonian is
\be
H=U\cdot H^d_0\cdot U^\dagger+diag(V_{eff},0,0),
\ee
with
\be
H^d_0=\frac{1}{2E_\nu}diag(-\delta m^2_{21},0,\delta m^2_{32}).
\ee
Here $V_{eff}$ is the effective potential (eq. \ref{MApotef}) and $U$ is the three neutrino  mixing matrix given by \citep{gon03,akh04,gon08,gon11},
\be
U =
{\pmatrix
{
c_{13}c_{12}                    & s_{12}c_{13}                    & s_{13}\cr
-s_{12}c_{23}-s_{23}s_{13}c_{12} & c_{23}c_{12}-s_{23}s_{13}s_{12}   & s_{23}c_{13}\cr
s_{23}s_{12}-s_{13}c_{23}c_{12}  &-s_{23}c_{12}-s_{13}s_{12}c_{23}   &  c_{23}c_{13}\cr
}},
\ee
where $s_{ij}=\sin\theta_{ij}$ and  $c_{ij}=\cos\theta_{ij}$. For anti-neutrinos one has to replace $U$ by $U^*$.  The different neutrino probabilities are given as
\bary
P_{ee}&=&1-4s^2_{13,m}c^2_{13,m}S_{31},\nonumber\\
P_{\mu\mu}&=&1-4s^2_{13,m}c^2_{13,m}s^4_{23}S_{31}-4s^2_{13,m}s^2_{23}c^2_{23}S_{21}-4
c^2_{13,m}s^2_{23}c^2_{23}S_{32},\nonumber\\
P_{\tau\tau}&=&1-4s^2_{13,m}c^2_{13,m}c^4_{23}S_{31}-4s^2_{13,m}s^2_{23}c^2_{23}S_{21}-4
c^2_{13,m}s^2_{23}c^2_{23}S_{32},\nonumber\\
P_{e\mu}&=&4s^2_{13,m}c^2_{13,m}s^2_{23}S_{31},\nonumber\\
P_{e\tau}&=&4s^2_{13,m}c^2_{13,m}c^2_{23}S_{31}\nonumber\\
P_{\mu\tau}&=&-4s^2_{13,m}c^2_{13,m}s^2_{23}c^2_{23}S_{31}+4s^2_{13,m}s^2_{23}c^2_{23}S_{21}+4
c^2_{13,m}s^2_{23}c^2_{23}S_{32},\nonumber\\
\label{prob}
\eary
where
\be
\sin
2\theta_{13,m}=\frac{\sin2\theta_{13}}{\sqrt{(\cos2\theta_{13}-2E_{\nu}V_{eff}/\delta
    m^2_{32})^2+(\sin2\theta_{13})^2}},
\ee
and
\be
S_{ij}=\sin^2\biggl(\frac{\Delta\mu^2_{ij}}{4E_{\nu}}l_{osc}\biggr)
\ee
with
\bary
\Delta\mu^2_{21}&=&\frac{\delta
  m^2_{32}}{2}\biggl(\frac{\sin2\theta_{13}}{\sin2\theta_{13,m}}-1\biggr)-E_{\nu}V_{eff}\nonumber\\
\Delta\mu^2_{32}&=&\frac{\delta
  m^2_{32}}{2}\biggl(\frac{\sin2\theta_{13}}{\sin2\theta_{13,m}}+1\biggr)+E_{\nu}V_{eff}\nonumber\\
\Delta\mu^2_{31}&=&\delta m^2_{32} \frac{\sin2\theta_{13}}{\sin2\theta_{13,m}}\,,
\eary
and
\bary
\sin^2\theta_{13,m}&=&\frac12\biggl(1-\sqrt{1-\sin^22\theta_{13,m}}\biggr)\nonumber\\
\cos^2\theta_{13,m}&=&\frac12\biggl(1+\sqrt{1-\sin^22\theta_{13,m}}\biggr).
\eary
The oscillation length for the neutrino is given by
\be
l_{osc}=\frac{l_v}{\sqrt{\cos^2 2\theta_{13} (1-\frac{2 E_{\nu} V_e}{\delta m^2_{32} \cos 2\theta_{13}}
    )^2+\sin^2 2\theta_{13}}},
\label{osclength}
\ee
where $l_v=4\pi E_{\nu}/\delta m^2_{32}$ is the vacuum oscillation length. The resonance condition and resonance length are,
\be\label{reso3}
V_{eff}-5\times 10^{-7}\frac{\delta m^2_{32,eV}}{E_{\nu,MeV}}\,\cos2\theta_{13}=0
\ee
and
\be
l_{res}=\frac{l_v}{\sin 2\theta_{13}}.
\ee
Considering the adiabatic condition at the resonance, we  can express it as, 
\bary
\kappa_{res}&\equiv & \frac{2}{\pi}
\left ( \frac{\delta m^2_{32}}{2 E_\nu} \sin 2\theta_{13}\right )^2
\left (\frac{dV_{eff}}{dr}\right)^{-1} \ge 1\nonumber\\
&=& 3.62\times 10^{-2} {\rm cm^{-1}}\left ( \frac{\delta m^2_{32,eV}}{E_{\nu,MeV}} \sin 2\theta_{13}\right )^2 \frac{1}{V^\prime}\ge 1,
\label{adbcon}
\eary
where
\be
V^\prime=\Omega_B\,\left[\frac{dV_{eff,1}}{dr} - 3.16\times 10^{-10} E_{\nu,MeV} \frac{dV_{eff,2}}{dr}\right].
\ee
and
\bary
V_{eff,1}&=& \sum^\infty_{l=0} (-1)^l \sin\alpha_l\,\biggl\{ \frac{m_e^2}{m^2_W}\biggl(1+4\frac{E^2_\nu}{m^2_e}\biggr)K_1(\sigma_l)\cr
&&+ \sum^\infty_{n=1} \lambda_n \left[2 + \frac{m_e^2}{m^2_W}\biggl(3+4\frac{E^2_\nu}{m^2_e} -2\Omega_B\biggr)\right]  K_1(\sigma_l\lambda_n)  \biggr\} \cr
V_{eff,2}&=&\sum^\infty_{l=0} (-1)^l \cos\alpha_l\left( \frac34K_0(\sigma_l)+   \sum^\infty_{n=1} \lambda^2_n\, K_0(\sigma_l\lambda_n)\right)
\eary
We will use the following parameters \citep{aha11,wen10} for this analysis:
\bary\label{3parosc}
{\rm for}&&\,\,\sin^2_{13} < 0.053: \delta m_{21}^2= (7.41^{+0.21}_{-0.19})\times 10^{-5}\,{\rm eV^2}\, {\rm and}\, \tan^2\theta_{12}=0.446^{+0.030}_{-0.029} \cr
{\rm for}&&\,\,\sin^2_{13} < 0.04: \delta m_{23}^2=(2.1^{+0.5}_{-0.2})\times 10^{-3}\,{\rm eV^2}\, {\rm and}\, \sin^2\theta_{23}=0.50^{+0.083}_{-0.093}
\eary
For a complete description of resonant neutrino oscillations at early fireball stage,  we use the values of  fireball observables during the initial stage and phase of acceleration.  At the initial stage, we consider a  fireball  endowed with magnetic field $B=10\,B_c$, temperature in the range 1 to 5 MeV and thermal neutrino energies  E$_\nu=$1,  5,  20  and  30 MeV whereas at the acceleration phase the corresponding values considered are:  magnetic field  $B=10^{-4.3}\,B_c$,  temperature in the range 50 to 500 keV and quasi-thermal GeV neutrino energies  E$_\nu=$1,  10,  20 and 50 GeV.  In both cases we take the best fit of parameters for two- (solar:  $\delta m^2=5.6 \times 10^{-5}\,{\rm eV^2}$ and  $\tan^2\theta=0.427$ \citep{aha11}, atmospheric: $\delta m^2=2.1\times 10^{-3}\,{\rm eV^2}$ and $\sin^22\theta=1.0$ \citep{abe11a},  and accelerator: $\delta m^2 \sim 0.6\, {\rm eV^2} $ and $\sin^22\theta = 51.2\times 10^{-3}$  \citep{chu02}) and  three ($\delta m^2_{32}= 10^{-2.58}$  and  $\theta_{13} =11^\circ $) -neutrino mixing in order to analyze the resonance conditions in both phases. \\ 
From the resonance conditions (eqs. \ref{reso2} and \ref{reso3}),  we have obtained the contour plots of temperature and chemical potential at the initial stage (fig. \ref{res_ab}) and phase of acceleration (fig. \ref{res_be}).  At the initial phase,  chemical potential lies in the range 7.4$\times 10^{-3}$ to 3.01 eV for solar parameters, 1.01$\times 10^{-2}$ to 2.84 eV for atmospheric parameters, 0.12 to 3.1 $\times 10^2$ keV for accelerator parameters and  0.8 to 50 $\times 10^2$ eV for three-neutrino mixing. One can see that temperature is a decreasing function of chemical potential which gradually increases as neutrino energy is decreased. In addition, we have computed  the resonance lengths which are shown in table 1.  As shown in this table, the resonance lengths lie in the range $l_{res}\sim$ (10$^4$ to $10^8$) cm which are comparable with the length scale of a fireball. Therefore,  depending on the  oscillation parameters  neutrinos  could oscillate resonantly before leaving the fireball. For instance, considering an initial radius of 10 km, only  neutrinos with low energy will oscillate resonantly  for atmospheric, accelerator and three-neutrino mixing but no solar parameters. If we assume an initial radius of 100 km, and  consider atmospheric and accelerator oscillation parameters,  then neutrinos would oscillate resonantly regardless their energies. Once again,  assuming a radius of 1000 km, thermal neutrinos will oscillate resonantly. 
\begin{table}
\begin{center}\renewcommand{\tabcolsep}{0.2cm}
\renewcommand{\arraystretch}{0.89}
\begin{tabular}{ccrlc} \hline
Energy                      &                                                      & $l_{res}$& (cm)& \\
{\small (MeV)}           &  {\small Solar}                               & {\small Atmosph.}   &  {\small Accelerat.}    & {\small Three flavors} \\\hline
{\small 1}                   &   {\small 4.8 $\times 10^6$}          & {\small 1.2 $\times 10^5$}                &  {\small 7.1 $\times 10^3$}     & {\small 2.5 $\times 10^5$ }    \\
{\small 5 }                  &    {\small 2.4 $\times 10^7$}          &  {\small 5.9 $\times 10^5 $}             &  {\small 3.6 $\times 10^4$ }    & {\small 1.3 $\times 10^6$}     \\
{\small 20}                 &    {\small 1.0 $\times 10^8$}          &   {\small  2.4 $\times 10^6$}            &  {\small 1.4 $\times 10^5$ }    &  {\small 5.0 $\times 10^7$}    \\
{\small 30}                 &    {\small  1.5 $\times 10^8$}         &   {\small  3.5 $\times 10^6$}            &  {\small 2.1 $\times 10^5$ }    & {\small  7.6 $\times 10^7$}  \\\hline
\end{tabular}
\end{center}
\caption{\small\sf Resonance lengths  of thermal neutrinos for the  best fit parameters of two- and three-neutrino mixing.}
\label{res_len1}
\end{table}
In the phase of acceleration,  chemical potential lies in the range 2.2$\times 10^{-2}$ to 1 eV for solar, 4.01$\times 10^{-2}$ to 1.3 eV for atmospheric, 1.1$\times10^{-2}$ to 0.5 keV for accelerator  and  0.5 to  6.1 eV for three-neutrino mixing. One can see that temperature is a decreasing function of chemical potential which increases as neutrino energy increases (decreases) for solar and atmospheric (accelerator) oscillation parameters  and it is doubly degenerate for  three-neutrino mixing.  Further, we have computed  the resonance lengths of GeV neutrinos.  As shown in table 2, the resonance lengths match  the length scale where they were created  for three-neutrino mixing. In other words, taking into account the former parameters, multi GeV neutrinos created at $\sim (10^{11}\,-\, 10^{13})$ cm  will  have  simultaneously resonant oscillations.  
\begin{table}
\begin{center}\renewcommand{\tabcolsep}{0.2cm}
\renewcommand{\arraystretch}{0.89}
\begin{tabular}{ccrlc} \hline
Energy                      &                                                      & $l_{res}$& (cm)& \\
{\small (GeV)}           &  {\small Solar}                               & {\small Atmosph.}   &  {\small Accelerat.}    & {\small Three flavors} \\\hline
{\small 5}                   &   {\small  2.4 $\times 10^{10}$}     &  {\small 5.9 $\times 10^8$}                 &  {\small 3.6 $\times 10^7$}   & {\small 1.3 $\times 10^9$}\\
{\small 10 }                &    {\small 4.8 $\times 10^{10}$}    & {\small 1.2 $\times 10^9$}                      &  {\small 7.1 $\times 10^7$}  & {\small 2.5 $\times 10^9$}\\
{\small 20}                 &    {\small 9.6 $\times 10^{10}$}     & {\small 2.4 $\times 10^9$}                      &  {\small 1.4 $\times 10^8$}    &  {\small 5.0 $\times 10^9$}\\
{\small 50}                 &    {\small 2.4 $\times 10^{11}$}      & {\small 5.9 $\times 10^9$} &  {\small  3.6 $\times 10^8$}    & {\small 1.3 $\times 10^{10}$}  \\\hline
\end{tabular}
\end{center}
\caption{\small\sf Resonance lengths  of quasi-thermal neutrinos for the  best fit parameters of the two- and three-neutrino mixing.}
\label{res_len2}
\end{table}
Finally, we have studied the survival and conversion probability for the active-active ($\nu_{e,\mu,\tau} \leftrightarrow \nu_{e,\mu,\tau}$) neutrino oscillations and the three-neutrino mixing.  In fig. \ref{prob} the survival probability of electron $P_{ee}$, muon  $P_{\mu\mu}$ and  tau  $P_{\tau\tau}$ neutrino and conversion probabilities  $P_{e\mu}$,  $P_{e\tau}$  and   $P_{\mu\tau}$   as a function of energy for fixed values of  length scale, temperature, magnetic  field and neutrino energy is plotted.  In the left-hand figures, we use once again the fireball values  in the initial stage: T= 5 MeV, r$_0$=100 km (above) and $r_0$=10 km (below), B=10 $B_c$, neutrino energy range 1 to 30 MeV and in the right-hand figures we take into account the values already mentioned for the phase of acceleration: T= 100 keV, r=10$^{11}$ cm (above) and r=10$^{12}$ cm (below), B= $10^{-4.3}\,B_c$, neutrino energy range 1- 30 GeV.  Our analysis shows that $P_{ee}=1$, $P_{e\mu}=P_{e\tau}=0$, indicating that the propagating electron neutrinos   do not oscillate to any other flavor  independently of the their  energies and  radii. Instead, it tells us that muon and tau neutrinos oscillate among themselves with equal probability and that the oscillation depends on their energy and distances.   As can be seen, the probabilities satisfy the condition 
\bary
\sum_{i=e,\mu,\tau} P_{ei} (\delta m^2_{32}, L)=1,\hspace{0.2cm} \sum_{i=e,\mu,\tau} P_{\mu i} (\delta m^2_{32}, L)=1,\hspace{0.2cm} \sum_{i=e,\mu,\tau} P_{\tau i} (\delta m^2_{32}, L)=1.\hspace{1cm}
\eary
\subsection{Neutrino Oscillation from  Source to Earth}
Between the surface of the star and the Earth the flavor ratio $\phi^0_{\nu_e}:\phi^0_{\nu_\mu}:\phi^0_{\nu_\tau}$   is affected by the full three description  flavor mixing, which is calculated as follows. The probability for a neutrino to oscillate from a flavor state $\alpha$ to a flavor state $\beta$ in a time starting from the emission of neutrino at star t=0, is given as
\bary
P_{\nu_\alpha\to\nu_\beta} &=&\mid <  \nu_\beta(t) | \nu_\alpha(t=0) >  \mid\cr
&=&\delta_{\alpha\beta}-4 \sum_{j>i}\,U_{\alpha i}U_{\beta i}U_{\alpha j}U_{\beta i}\,\sin^2\biggl(\frac{\delta m^2_{ij}\, l_{osc}}{4\, E_\nu}   \biggr)\,.
\eary
Using the set of parameters given in eq. (\ref{3parosc}), we can write the mixing matrix
\be
U =
{\pmatrix
{
0.816669	   &  0.544650     &     0.190809\cr
 -0.504583  & 0.513419      &	 0.694115\cr
 0.280085   &  -0.663141    &     0.694115\cr
}}\,.
\ee
Additionally, averaging the sin term in the probability to $\sim 0.5$ for larger distances $l_{osc}$ \citep{lea95}, the probability matrix for a neutrino flavor vector of ($\nu_e$, $\nu_\mu$, $\nu_\tau$)$_{source}$ changing to a flavor vector  ($\nu_e$, $\nu_\mu$, $\nu_\tau$)$_{Earth}$ is given as
\be
{\pmatrix
{
\nu_e   \cr
\nu_\mu   \cr
\nu_\tau   \cr
}_{Earth}}
=
{\pmatrix
{
0.534143	  & 0.265544	  & 0.200313\cr
 0.265544	  & 0.366436	  &  0.368020\cr
 0.200313	  & 0.368020	  & 0.431667\cr
}}
{\pmatrix
{
\nu_e   \cr
\nu_\mu   \cr
\nu_\tau   \cr
}_{source}}\,,
\label{matrixosc}
\ee
for distances longer than the solar system. 
\vspace{3cm}
\section{Dynamics of  the Fireball}
Although we have already introduced general concepts of the fireball model and its dynamics, and  also calculated the range of values of chemical potentials and temperatures for which the resonance condition is satisfied,  here we are going to quantify or/and estimate  the observable quantities in the evolution of the fireball requiring the charged-neutrality condition in addition to resonance condition.  In this manner,  we are going to constrain the dynamics of the fireball  by  means of neutrino oscillations.\\
\subsection{Initial stage}
The initial state of the fireball is magnetized,  hot  and dense of  baryons  and pairs  $e^{\pm}$ in perfect thermodynamic equilibrium  with a comoving temperature
\be
T'=\biggl(\frac{L_{52}}{4\pi r^2_{0,6} a}\biggr)^{1/4}  \simeq 3.8\, {\rm MeV}\,,
\ee
where $a=\pi^2\,k^4/15=7.6\times 10^{-15}\, {\rm erg\,cm^{-3}\,K^{-4}}$ is the radiation density constant, L is the total isotropic-equivalent energy outflow in the jet and $r_0$ is the initial radius.  Magnetic fields could be estimated  by means of the equipartition of the total energy density (U)
\be
B\simeq \sqrt{8\pi\epsilon_B U},
\ee
which becomes  important as the energy density is dominated by baryons \citep{bel03a} 
\be
U\simeq \frac32 \frac{T\rho}{m_p}\,,
\ee
and the electrons are degenerated with the charged-neutrality condition, $n_e(\mu, T, B) - \bar{n}_e(\mu, T, B)=n_p$ where $n_e(\mu, T, B)$ and $\bar{n}_e(\mu, T, B)$ are the number densities of electrons and positrons generated in the plasma  a temperature T and magnetic field B, and n$_p$ is the number density of protons.   Taking into account that the number density of neutrons is comparable to that of protons and from the number density of electrons and positrons at the moderate magnetic field and relativistic temperatures found in section II  (eq. \ref{B1}), the changed-neutrality condition is given by
\be\label{cha_con}
\frac{m^3_e}{\pi^2}\Omega_B \sum^\infty_{l=0}(-1)^l \sinh\alpha_l\,\biggl\{ K_1(\sigma_l)+ 2 \sum^\infty_{n=1}\lambda_n\,K_1(\sigma_l\lambda_n)\biggr\}=Y_e\frac{\rho}{m_p}\,,
\ee
here $\sigma_l$ and $\alpha_l$ correspond to the values found for which resonance condition is satisfied,  $\rho$ is the baryon density  and $Y_e=n_p/(n_n+n_p)$ is the proton-to-nucleon ratio.  The baryon density is defined by means of the total mass outflow rate in baryon loaded jet  $\dot{M}=4\pi r_0^2\rho$ and the dimensionless entropy or  baryon load parameter $\eta=L/\dot{M}$. The entropy per baryon is conserved in the flow and related to $\eta$ and T as $s/k_B=4\eta m_p/3T$. Initially, due to high temperatures $T\gg m_e$ and charged current reactions   
\be
e^-+p\to n+\nu, \hspace{2cm}e^++n\to p+\bar{\nu},
\ee
protons convert into neutrons and neutrons into protons, achieving  a balance
\be
 Y_e=\frac12+\frac{7\pi^4}{1350\,\zeta(5)}\biggl(\frac{Q/2-\mu}{T}\biggr) 
\ee
between the rates of $e^-$ and $e^+$ capture by $n_p$ and $n_n$ with Q defined through neutron ($m_n$) and proton ($m_p$) mass, $Q=m_n - m_p$ \citep{bel03a}. Hence, the changed-neutrality condition can be written as
\be\label{cha_con1}
\sum^\infty_{l=0}(-1)^l \sinh\alpha_l\,\biggl\{ K_1(\sigma_l)+ 2 \sum^\infty_{n=1}\lambda_n\,K_1(\sigma_l\lambda_n)\biggr\}=\frac{\rho\,\Omega^{-1}_B}{m_p\,m^3_e} \biggl[\frac{\pi^2}{2}+\frac{7\pi^6}{1350\,\zeta(5)}\biggl(\frac{Q/2-\mu}{T}\biggr)\biggr].
\ee
It is very important to highlight that the proton-to-nucleon ratio is a function of temperature and chemical potential of the initial stage of the fireball, when the initial optical depth is extremely high and the  phase of acceleration has not yet begun. \\
Taking into account the range of values of chemical potentials and temperatures  found from resonance conditions (figs. \ref{res_ab} and \ref{res_be}), considering  magnetic fields at the moderate field limit, 50 $B_c$ (left-hand figures) and  0.1 $B_c$ (right-hand figures) and requiring the charged-neutrality condition (eqs. \ref{cha_con} and \ref{cha_con1}),  we plot contour lines  of baryon density ($\rho$) (above) and the proton-to-nucleon ratio (Y$_e$) (below)  as a function of  temperature as shown in fig. \ref{ye}.  From the figures above, one can see that the highest line of  baryon density  increases gradually  achieving a maximum density  of 7.2 $\times 10^7$ g cm$^{-3}$ (left-hand figure) and 5.6 $\times 10^6$ g cm$^{-3}$ (right-hand figure) for  2 MeV $\leq T \leq$ 3 MeV and $\mu=160$ keV. This value of chemical potential is the largest one obtained from the resonance condition,  hence  resonant oscillations  are suppressed for densities larger than these. From the figures below one can see that proton-to-nucleon ratio is a decreasing function of temperature achieving a minimum value of $\sim$ 0.52 and 0.53 for baryon density  $\simeq$10$^8$ g cm$^{-3}$ (left-hand figure)  and  $\simeq$10$^6$ g cm$^{-3}$  (right-hand figure) and maximum value of $\sim$ 0.87 and  0.88 for baryon density 10$^6$ g cm$^{-3}$ (left-hand figure)  and  10$^{4.5}$ g cm$^{-3}$,  therefore resonant oscillations are only allowed  when the number density of protons is at least slightly larger that that of neutrons $n_p \geq n_n$.  This result could make evident the de-neutralization process in the fireball,  where $Y_e$ is less than 0.5 at early times (there are not resonant oscillations) and after it becomes larger than 0.5 ($Y_e > 0.5$) at latter times (resonant oscillations are allowed).    Other important characteristic from the figures below is that  the number density of neutrons to protons ratio n$_n$/n$_p$ = (1-Y$_e$)/Y$_e$ is lower for that fireball with more magnetization. For instance,  taking the value of density $\rho$=10$^6$ g cm$^{-3}$ in both plots, we can see that $Y_e$(n$_n$/n$_p$) = 0.865(0.156) for B=50 B$_c$ and $Y_e$(n$_n$/n$_p$) = 0.751(0.332) for B=0.1$B_c$.\\
\subsection{Phase of acceleration}
When fireball starts expanding, the optically thick hot plasma expands with an increasing bulk Lorentz factor $\Gamma\propto r/r_0$ with radius $r$ following the adiabatic law and the comoving temperature drops as $T'(r)\propto r_0/r$.  Protons and neutrons keep coupled with each other until the value of Lorentz factor $\Gamma\simeq  \eta$ is larger or less than a critical value or  the dynamical time $t'_{np}\simeq (r/\Gamma)$ is shorter than the elastic scattering  time $t'_{np}\simeq (n'_p\,\sigma_{np})^{-1}$ where $\sigma_{np}\approx 3\times 10^{-26}$ cm$^2$ and $n'_p$ is the number density of protons.  The critical value is defined by 
\be\label{c_lorentz} 
\eta_\nu=\left( \frac{L\, \sigma_{np}\,Y_e}{4\pi\,m_p\,r_0}\right)^{1/4}\simeq4.6\times 10^2L^{1/4}_{52}\,r_{o7}^{-1/4}\,Y_e^{1/4}  \,,
\ee
and depending on both $\eta$ and $\eta_\nu$ different processes of decoupling in a p-n outflow take place leading to  varied energy ranges of neutrinos. For $\eta\geq \eta_\nu$, neutrinos  in the energy range  5-10 GeV  are expected \citep{bac00} whereas $\eta\leq \eta_\nu$  neutrino energy  lies in the range  2-25 GeV or higher depending on the value of $\eta$ \citep{mes00}.   Supposing that the decoupling takes place  before coasting, then a slowly hadron shell ($\Gamma_s$) is overtaken by another  shell  that in principle moves faster ($\Gamma_f=\eta$) in the flow, producing inelastic collisions at r$\sim 2 \Gamma^2_s\,r_0$. Following \citet{mes00},  the typical collision Lorentz factor can be written as $\Gamma_{rel}\simeq 1/2(\Gamma/\Gamma_s + \Gamma_s/\Gamma)$ with $\Gamma=\sqrt{\Gamma_s\Gamma_f}\simeq\Gamma_f$, and the total energy is $E_{Tot}=(2m^2_p+2m_pE_{rel})^{1/2}$, whereas the CM threshold energy is $2m_p+m_\pi\sim 2$ GeV. Hence, at  depths  $\tau_{pn}\sim n'_p\sigma_{pn} (r/\Gamma) > 1$ each neutron heated by an individual shock could collide $k_\pi \sim 2$ times before its CM energy becomes less than threshold.   From the total number of neutrons involved in the shock $N_n=(1-Y_e) \epsilon E/(\eta\,m_p)$  and assuming  a shock dissipation efficiency $\epsilon=0.2$,  the number of expected events per year in DeepCore detector is
\be\label{n_neu}
R_{\nu\bar{\nu}}\sim 0.33 E_{53}(1-Y_e)\,N_{t37.7}\,R_{b3}\,h^2_{65}\,\left(\frac{2-\sqrt2}{1+z-\sqrt{1+z}}\right)^2\,{\rm year}^{-1}\,,
\ee
where we have taken into account  the product of density of ice and the effective volume at 100 GeV as $\rho_{ice}\,V_{eff}\simeq$ 50 megaton ($V_{eff}\simeq\, 5.56 \times 10^{-2}\, {\rm km}^3$ and  target protons of N$_t\sim 10^{37.7} N_{t37.7}$) \citep{abb12a}, a burst rate out to a Hubble radius of $10^3$ R$_b$ and an Einstein-de-Sitter universe with Hubble constant H=65 h$_{65}$ km/s/Mpc.   Recently, \citet{koh13} showed that for a low-luminosity GRBs at D=10 Mpc with $\Gamma$=10 and neutron luminosity L$_n=2\times10^{46}$ erg/sec, quasi-thermal neutrinos around  10 GeV are expected in DeepCore.\\ 
We plot the  critical Lorentz factors ($\eta_\nu$) (figure above) and number of neutrinos ($R_{\nu\nu}$) (figure below) as a function of  temperature when the resonance and charged-neutrality conditions are satisfied as shown in fig. {\ref{neut}.  From figures above,  one can see that the critical Lorentz factor  is a decreasing function of temperature and decreasing function of baryon density.  Besides,  $\eta_\nu$ lies in the range  $\simeq$ 436 to 662 for  temperatures and baryon densities in the range 50 to 10$^3$ keV and 10 to 10$^3$ (1 to 10$^2$) g/cm$^3$ for B= $10^{-4.5}\, B_c$ (B= $10^{-5.5}\,B_c$). From these plots can be seen that the effect of magnetic field is to increase $\eta_\nu$. In figures below, we can observe that the number of expected neutrinos is an increasing function of temperature and  baryon density.  Comparing both plots we see that as  $\rho= 10^2$  g/cm$^3$ and T= 0.7 MeV the expected neutrinos are $\simeq$ 2.23  for B= $10^{-4.3} B_c$ and $\simeq$ 4 for B= $10^{-5.3} B_c$, which tells us that the effect of magnetic field is to decrease the number of expected neutrinos.  From that we can see the importance of knowing the strength of magnetic field and as it could alter the dynamics.\\ 
It is important to notice that independently of the model and considerations assumed \citep{bac00,mes00,koh13} multi-GeV neutrinos are expected on Earth, hence we will estimate their flavor ratio.   Considering the flux ratio for $\pi^\pm$ and $\mu^\pm$ decay  as $\dot{N}_{\nu_\mu}\simeq\dot{N}_{\bar{\nu}_\mu}\simeq 2 \dot{N}_{\nu_e}\simeq2\dot{N}_{\bar{\nu}_e}$\citep{raz06a}  and  from oscillation probabilities at 10$^{11}$ cm, 10$^{12}$ cm  and 10$^{13}$ cm given in the section III, we compute the flavor ratio  for neutrino energies of 5, 10, 20 and 50 GeV as shown  in table \ref{flaratio_s}.  From this table one can see an interesting result, that although tau neutrino at GeV energies is not created by p-n decoupling, it appears due to the resonant oscillations of muon neutrinos in the fireball.   
\begin{table}
\begin{center}\renewcommand{\tabcolsep}{0.2cm}
\renewcommand{\arraystretch}{0.89}
\begin{tabular}{|c|c|c|c|c|c|}\hline
$E_{\nu}$  &$\phi_{\nu_e}:\phi_{\nu_\mu}:\phi_{\nu_\tau}$ &$\phi_{\nu_e}:\phi_{\nu_\mu}:\phi_{\nu_\tau}$ &$\phi_{\nu_e}:\phi_{\nu_\mu}:\phi_{\nu_\tau}$\\
(GeV)&(r=10$^{11}$ cm)&(r=10$^{12}$ cm)&(r=10$^{13}$ cm)\\ \hline

5       &  1.000:1.959:0.041  &  1.000:0.021:1.979 &  1.000:0.848:1.152  \\\hline

10     & 1.000:0.164:1.837  &  1.000:0.951:1.049&  1.000:0.000:2.000\\\hline

20     & 1.000:1.101:0.899  & 1.000:0.282:1.718 &  1.000:1.917:0.083\\\hline

50     & 1.000:0.675:1.325  &  1.000:0.002:1.998   &  1.000:1.839:0.162\\\hline

\end{tabular}
\end{center}
\caption{\small\sf The flavor ratio on the surface of the fireball  for four neutrino energies  (E$_{\nu}$=5 GeV, 10 GeV, 20 GeV and 50 GeV), leaving the star to three distances r=10$^{11}$ cm, 10$^{12}$ cm and 10$^{13}$ cm.}
\label{flaratio_s}
\end{table}
In addition, from table \ref{flaratio_s}  and eq. \ref{matrixosc}  we estimate the flavor ratio expected on  Earth  for the same range of neutrino energy as shown in  table \ref{flaratio_e}.
\begin{table}
\begin{center}\renewcommand{\tabcolsep}{0.2cm}
\renewcommand{\arraystretch}{0.89}
\begin{tabular}{|c|c|c|c|c|c|}\hline
$E_{\nu}$  &$\phi^0_{\nu_e}:\phi^0_{\nu_\mu}:\phi^0_{\nu_\tau}$ &$\phi^0_{\nu_e}:\phi^0_{\nu_\mu}:\phi^0_{\nu_\tau}$ &$\phi^0_{\nu_e}:\phi^0_{\nu_\mu}:\phi^0_{\nu_\tau}$ \\
(TeV)&(r=10$^{11}$ cm)&(r=10$^{12}$ cm) &(r=10$^{13}$ cm) \\ \hline
5      &  1.063:0.998:0.939   &  0.936:1.002:1.062&  0.990:1.000:1.010 \\\hline
10    &  0.945:1.001:1.053  &  0.997:1.000:1.003 &  0.935:1.002:1.064\\\hline
20    &   1.007:0.999:0.994 &  0.953:1.001:1.046  & 1.060:0.999:0.941\\\hline
50    &  0.979:1.001:1.021  &  0.935:1.002:1.064 &  1.055:0.999:0.947\\\hline
\end{tabular}
\end{center}
\caption{\small\sf The flavor ratio expected on Earth for four neutrino energies  (E$_{\nu}$=5 GeV, 10 GeV, 20 GeV and 50 GeV), leaving the star to three distances r=10$^{11}$ cm, 10$^{12}$ cm and 10$^{13}$ cm.}
\label{flaratio_e}
\end{table}
As shown in this table, GeV sub-photospheric neutrinos  with deviations of this standard flavor ratios are expected.
\section{Results and Conclusions}
We have explicitly calculated the neutrino self-energy  and neutrino effective potential up to an order $m^{-4}_W$ as a function of temperature, chemical potential, magnetic field and neutrino energy. We have shown that for neutrinos in the GeV energy range  as well as  small chemical potentials which is the case of solar and atmospheric neutrino parameter (small particle-antiparticle asymmetry)  the contribution of m$^{-4}_W$ terms to the neutrino effective potential is relevant.  In the magnetic field framework, we have derived it at the strong, moderate (above and below B$_c$)  and weak-field limit,  taking into account a background composed of  pairs $e^\pm$, photons, protons, neutrons and neutrinos.  Also we have derived the resonance and  charged-neutrality conditions.   By considering the neutrino effective potential at the moderate field limit, using the typical values of a magnetized GRB fireball and requiring the resonance and  charged-neutrality conditions,  we have studied the thermal and quasi-thermal neutrino oscillations  assuming a neutron abundance which is comparable to that of protons. In the fireball scenario,  thermal neutrino oscillations have been studied at the initial stage ($r_0\simeq 10^{6.5} -10^{7.5}\, {\rm cm},\, B\simeq 0.1 - 50\, B_c$ and $T\simeq 1 - 10$) MeV whereas  quasi-thermal GeV neutrinos in the phase of acceleration (r$\simeq 10^{11} -10^{13}\, {\rm cm},\, B\simeq 10^{-4.3} - 10^{-5.3}\, B_c$ and $T\simeq 50 - 700$ keV). This complete analysis has been carried out  using the two- (solar, atmospheric and accelerator neutrino parameters) and  three-neutrino mixing.\\
The results for the initial stage are:\\
\begin{enumerate}
\item Resonant oscillations are suppressed for baryon densities greater than $\sim\,10^8$ g cm$^{-3}$  for 50 $B_c$   and $\sim\,10^6$ g cm$^{-3}$  for 0.1 $B_c$.\\
\item  Neutrinos can hardly oscillate resonantly for a fireball with number density of neutrons greater than protons $n_n\geq n_p$ or for proton-to-baryon ratio larger than 0.52 ($Y_e\leq$ 0.52).\\
\item The number density of neutrons to protons ratio ($n_n/n_p$) is lower for that fireball with more magnetization.  It is due to positron capture on neutrons is greatly accelerated by the large magnetic phase space factor \citep{tho13}.  Also baryon densities are larger for a fireball with more magnetization. 
\end{enumerate}
From the phase of acceleration, we showed that:\\
\begin{enumerate}
\item The critical Lorentz factor for neutrino production  is limited by the load density, temperature  and the magnetic field.  The effect of magnetic field in the emission region is to decrease the expected number of neutrinos.\\
\item GeV neutrinos created in the sub-photospheric region can oscillate resonantly. Due to this, we estimate the neutrino flavor ratio and deviations of  standard flavor ratios are expected.    
\end{enumerate}
Neutrinos of energies of 1 to 30 MeV are very similar to those produced by type II supernova i.e, SN1987A, however are of cosmological distance.  These cosmological events make the thermal  neutrino flux very low on Earth compared to the ones we had seen from the supernova SN1987A.  Although with the current neutrino telescopes, low energy neutrinos would be pretty difficult to detect, such survey would help us to understand the dynamics of the jet as it changes with the content of baryons.
\acknowledgments
We thank the anonymous referee  for a critical reading of the paper and valuable suggestions that helped improve the quality and clarity of this work.  We also thank  Bing Zhang,  Kohta Murase,  Ignacio Taboada, William Lee, Dany Page and Enrique Moreno for useful discussions.  This work was supported by Luc Binette scholarship and the projects IG100414 and Conacyt 101958.

\clearpage

\appendix

\section{Strong Magnetic field: $\Omega_B \gg 1$ }
In the strong magnetic field approximation, the energy of charged particles is modified confining the particles to the lowest Landau level ($n=0$). Thus,  the number density of electrons given by Eq. (\ref{den}) will  become
\be
n_e^0=\frac{eB}{2\pi^2}\int^\infty_0 dp_3 f_{e,0}\,,
\ee
where 
\be
f(E_{e,0})=\frac{1}{e^{\beta(E_{e,0}-\mu)} +1}\,,
\ee
and the  electron energy in the lowest Landau level is,
\be
E^2_{e,0}=(p^2_3+m_e^2)\,.
\ee
Assuming that  the chemical potentials ($\mu$) of the electrons and positrons are smaller than their energies ($\mu\leq$E$_e$), the fermion distribution function can be written as a sum  given by
\be
f(E_{e,0})=\frac{1}{e^{\beta(E_{e,0}-\mu)} +1}\approx\sum^{\infty}_{l=0}(-1)^l e^{-\beta(E_{e,0}-\mu)(l+1)}\, .
\ee
Therefore, the Lorentz scalars in this approximation are reduced to \citep{eli04,sah09a}
\bary
b_W=&&\sqrt2 G_F\biggl[ \biggl( 1+\frac32\frac{m_e^2}{m_W^2}+ \frac{eB}{m_W^2} +\frac{E_{\nu_e}k_3}{m_W^2}+\frac{E^2_{\nu_e}}{m_W^2}\biggr)(N^0_e - \bar{N}^0_e)\cr
&& -\frac{eB}{2\pi^2m_W^2}\int^\infty_0\,dp_3\biggl\{2\,k_3E_{e,0}+2E_{\nu_e}\biggl( E_{e,0} - \frac{m_e^2}{2E_{e,0}}\biggr) \biggr\}  (f_{e,0}+\bar{f}_{e,0})       \biggr]
\label{Lescb}
\eary
and
\bary
c_W=&&\sqrt2 G_F\biggl[ \biggl( 1+\frac12\frac{m_e^2}{m_W^2}+ \frac{eB}{m_W^2} -\frac{E_{\nu_e}k_3}{m_W^2}-\frac{k^2_3}{m_W^2}\biggr)(N^0_e - \bar{N}^0_e)\cr
&& -\frac{eB}{2\pi^2m_W^2}\int^\infty_0\,dp_3\biggl\{ 2E_{\nu_e}\biggl( E_{e,0} - \frac{m_e^2}{2E_{e,0}}\biggr) \biggr\}      +2k_3\biggl( E_{e,0} - \frac{3m_e^2}{2E_{e,0}}\biggr) \biggr\}  (f_{e,0}+\bar{f}_{e,0})       \biggr]\,,
\label{Lescc}
\eary
For Z-exchange diagram we do not have magnetic contribution and for the tadpole diagram only electron loop will be affected by the magnetic field \citep{sah09b}. The electron number density and other useful quantities at the strong field limit are
\be
N^0_e=\frac{m^3}{2\pi^2}\frac{B}{B_c}\sum^\infty_{l=0}(-1)^le^\alpha_l\,K_1(\sigma_l),
\ee
\be
N^0_e - \bar{N}^0_e =\frac{B}{B_c}\frac{m_e^3}{\pi^2}\,\sum^\infty_{l=0} (-1)^l \sinh\alpha_l\,K_1(\sigma_l),
\label{NONO}
\ee
\bary
\frac{eB}{2\pi^2}\int^\infty_0\,dp_3\,E_0(f_{e,0}+\bar{f}_{e,0})=&&\frac{m_e^4}{\pi^2}\frac{B}{B_c}\sum^\infty_{l=0} (-1)^l \cosh\alpha_l\, \biggl(K_0(\sigma_l) +\frac{K_1(\sigma_l)}{\sigma_l}\biggl),\cr
\frac{eB}{2\pi^2}\int^\infty_0\,dp_3\,\frac{1}{E_0}(f_{e,0}+\bar{f}_{e,0})=&&\frac{m_e^4}{\pi^2}\frac{B}{B_c}\sum^\infty_{l=0} (-1)^l \cosh\alpha_l\,K_0(\sigma_l), 
\label{usefulS}
\eary
where we have defined
\be
\alpha_l=\beta\mu(l+1)\hspace{1cm}\rm{and} \hspace{1cm}\sigma_l=\beta m_e(l+1)
\ee
with K$_i$ is the modified Bessel function of integral order i.  Replacing Eqs. (\ref{NONO}) and (\ref{usefulS}) in the Lorentz scalars (Eqs. \ref{Lescb} and \ref{Lescc}),  we obtain 
\bary
b_W&=&\frac{\sqrt2 G_F\,m_e^3}{\pi^2}\frac{B}{B_c}\biggl[\biggl\{1+\frac{m^2_e}{m^2_W}\biggl(\frac32+2\frac{E^2_{\nu,e}}{m^2_e}  +\frac{B}{B_c}  \biggr)  \biggr\}  \sum^\infty_{l=0} (-1)^l \sinh\alpha_l\,K_1(\sigma_l)\cr
&&-\frac{m^2_e}{m^2_W}\frac{E_{\nu,e}}{m_e}  \sum^\infty_{l=0} (-1)^l \cosh\alpha_l\biggl\{3K_0(\sigma_l)+4\frac{K_1(\sigma_l)}{\sigma_l} \biggr\} \biggr]\,,
\label{bf}
\eary
\bary
c_W&=&\frac{\sqrt2 G_F\,m_e^3}{\pi^2}\frac{B}{B_c}\biggl[\biggl\{1+\frac{m^2_e}{m^2_W}\biggl(\frac12-2\frac{E^2_{\nu,e}}{m^2_e}  +\frac{B}{B_c}  \biggr)  \biggr\}  \sum^\infty_{l=0} (-1)^l \sinh\alpha_l\,K_1(\sigma_l)\cr
&&-4\frac{m^2_e}{m^2_W}\frac{E_{\nu,e}}{m_e}  \sum^\infty_{l=0} (-1)^l \cosh\alpha_l \frac{K_1(\sigma_l)}{\sigma_l}  \biggr]\,.
\label{cf}
\eary
Finally,  from  Eqs. (\ref{poteff}), (\ref{bf}) and  (\ref{cf}) and for neutrinos moving  along the direction of magnetic field,  the neutrino effective potential in the strong magnetic field regime is
\bary
V_{eff}=&&\frac{\sqrt2 G_F\,m_e^3}{\pi^2}\frac{B}{B_c}\biggl[\sum^\infty_{l=0} (-1)^l \sinh\alpha_l\,K_1(\sigma_l)\biggl\{ 1+ \frac{m_e^2}{m^2_W}\biggl(\frac32+2\frac{E^2_\nu}{m^2_e} +\frac{B}{B_c}  \biggr)\cr
&&-\biggl(1+ \frac{m_e^2}{m^2_W}\biggl(\frac12-2\frac{E^2_\nu}{m^2_e} +\frac{B}{B_c}  \biggr) \biggr)\cos\phi  \biggr\}\cr
&& -4\frac{m_e^2}{m^2_W}\frac{E_\nu}{m_e}\sum^\infty_{l=0} (-1)^l \cosh\alpha_l\biggl\{ \frac34K_0(\sigma_l)+\frac{K_1(\sigma_l)}{\sigma_l}-\frac{K_1(\sigma_l)}{\sigma_l}\cos\phi\biggr\}\biggr]\,.
\eary
Doing $\Omega_B=B/B_c$ and   the previous effective potential is written as shown in eq. (\ref{Spotef}).
\section{ Moderate Magnetic field: $\Omega_B > 1$ and $\Omega_B \leq 1$  }
The electron energy in the magnetic field is given by
\be
E_{e,n}^2=(p_3^2+m_e^2+2neB)=p_3^2+m_e^2(1+2n\Omega_B),
\ee
In this case,  the number density of electrons (eq. \ref{den}) is written as
\be
n_e=\frac{\Omega_B\,m_e^2}{2\pi^2}\biggl[ \int_0^\infty dp_3 f_{e,0} +2\,\sum_{n=1}^\infty \int_0^\infty dp_3 f_{e,n}\biggr]\,,
\ee
and the electron distribution function by
\be
f(E_{e,n})=\frac{1}{e^{\beta(E_{e,n}-\mu)} +1}\approx\sum^{\infty}_{l=0}(-1)^l e^{-\beta(E_{e,n}-\mu)(l+1)}\, .
\ee
Calculating the below useful quantities in this regime,
\be\label{B1}
n_e-\bar{n}_e=\frac{m^3}{\pi^2}\Omega_B\biggl[ \sum^\infty_{l=0}(-1)^l \sinh\alpha_l\,\biggl\{ K_1(\sigma_l)+ 2 \sum^\infty_{n=1}\lambda_n\,K_1(\sigma_l\lambda_n)\biggr\}\biggr]
\ee
\be\label{B2}
n_e+\bar{n}_e=\frac{m^3}{\pi^2}\Omega_B\biggl[ \sum^\infty_{l=0}(-1)^l \cosh\alpha_l\,\biggl\{ K_1(\sigma_l)+ 2 \sum^\infty_{n=1}\lambda_n\,K_1(\sigma_l\lambda_n)\biggr\}\biggr]
\ee
\be\label{B3}
\int^\infty_n\,dp_3\,E_n(f_{e,n}+\bar{f}_{e,0})= 2 m_e^2 \sum^\infty_{l=0} (-1)^l \cosh\alpha_l\, \biggl[ \biggl( K_0(\sigma_l) +\frac{K_1(\sigma_l)}{\sigma_l}\biggl) +2\sum^\infty_{n=1}  \lambda_n^2  \biggl( K_0(\sigma_l\lambda_n) +\frac{K_1(\sigma_l\lambda_n)}{\sigma_l}\biggl)\biggr] ,
\label{useful}
\ee
\be\label{B4}
\int^\infty_0\,dp_3\,\frac{1}{E_n}(f_{e,n}+\bar{f}_{e,n})=2\sum^\infty_{l=0} (-1)^l \cosh\alpha_l\,\biggl[K_0(\sigma_l)+\frac12\sum^\infty_{n=1}\,K_0(\sigma_l\lambda_n)\biggr] , 
\ee
where $\lambda$ is defined by 
\begin{equation}
\label{espic}
\lambda^2=
\cases{
2\,n\,\Omega_B& for $\Omega_B>$ 1\hspace{1cm}{\rm moderately above}\cr
1+2\,n\,\Omega_B &for  $\Omega_B\leq$ 1\hspace{1cm}{\rm moderately below}\,, \cr
}
\end{equation}
with the Lorentz scalars and  again with $A_e=\sqrt2 G_F\,\frac{m_e^3\,\Omega_B}{\pi^2}$, the neutrino effective potential in the moderate regimen can be written as shown in eq. (\ref{MApotef}).
\section{ Weak Magnetic field: $\Omega_B \ll 1$ }
In this regimen, sums over the parameter  $\lambda =\sqrt{1+2n\,\Omega_B}$ can be approximated  by an integral  $\sum_n\to\int {\small dn}$. Therefore, it is useful  to write the following relations
\bary
\sum^\infty_{n=0}\lambda_n^2 K_0(\sigma_l\lambda_n)&=&\Omega_B^{-1}\biggl[ 2\frac{K_0(\sigma_l)}{\sigma_l}+\biggl( 1+\frac{4}{\sigma_l^2}\biggr)\frac{K_1(\sigma_l)}{\sigma_l}   \biggr]\cr
\sum^\infty_{n=0}\lambda_n K_1(\sigma_l\lambda_n)&=&\Omega_B^{-1}\biggl[ \frac{K_0(\sigma_l)}{\sigma_l}+ \frac{2}{\sigma_l^2}\,K_1(\sigma_l)   \biggr]\,,
\eary
where we have applied the  integrals
\bary
\int^\infty_1\frac{t^2dt}{\sqrt{t^2-1}}e^{-\sigma_l t}&=&K_0(\sigma_l)+\frac{K_1(\sigma_l)}{\sigma_l}\cr
\int^\infty_1\frac{tdt}{\sqrt{t^2-1}}e^{-\sigma_l t}&=&K_1(\sigma_l)\cr
\int^\infty_1\frac{dt}{\sqrt{t^2-1}}e^{-\sigma_l t}&=&K_0(\sigma_l)
\eary
for  $Re\,\sigma_l > 0$ and we have used the recurrence relation  $K_3(\sigma_l)=\frac{4}{\sigma_l}K_0(\sigma_l)+ \biggl( 1+\frac{8}{\sigma_l^2}\biggr)\,K_1(\sigma_l)$ and $K_2(\sigma_l)=K_0(\sigma_l)+ \frac{2}{\sigma_l}\,K_1(\sigma_l)$. Therefore,  from eqs. (\ref{B1}), (\ref{B2}) and (\ref{MApotef}),  the neutrino effective potential is given by eq. (\ref{Wpotef}) and  useful functions of electron number density can be written as
\be
n_e-\bar{n}_e=\frac{m^3}{\pi^2}\Omega_B\biggl[ 2\sum^\infty_{l=0}(-1)^l \sinh\alpha_l\,   \biggl\{\biggl( \frac{K_0(\sigma_l)}{\sigma_l} +  \frac{K_1(\sigma_l)}{\sigma_l^2}\biggr)\Omega^{-1}_B-\frac12 K_1(\sigma_l)\biggr\}\biggr]
\ee
\be
n_e+\bar{n}_e=\frac{m^3}{\pi^2}\Omega_B\biggl[ 2\sum^\infty_{l=0}(-1)^l \cosh\alpha_l\, \biggl\{\biggl( \frac{K_0(\sigma_l)}{\sigma_l} +  \frac{K_1(\sigma_l)}{\sigma_l^2}\biggr)\Omega^{-1}_B-\frac12 K_1(\sigma_l)\biggr\}\biggr]\,.
\ee
\clearpage

\begin{figure*}[htp]
\centering
\includegraphics[width=\textwidth]{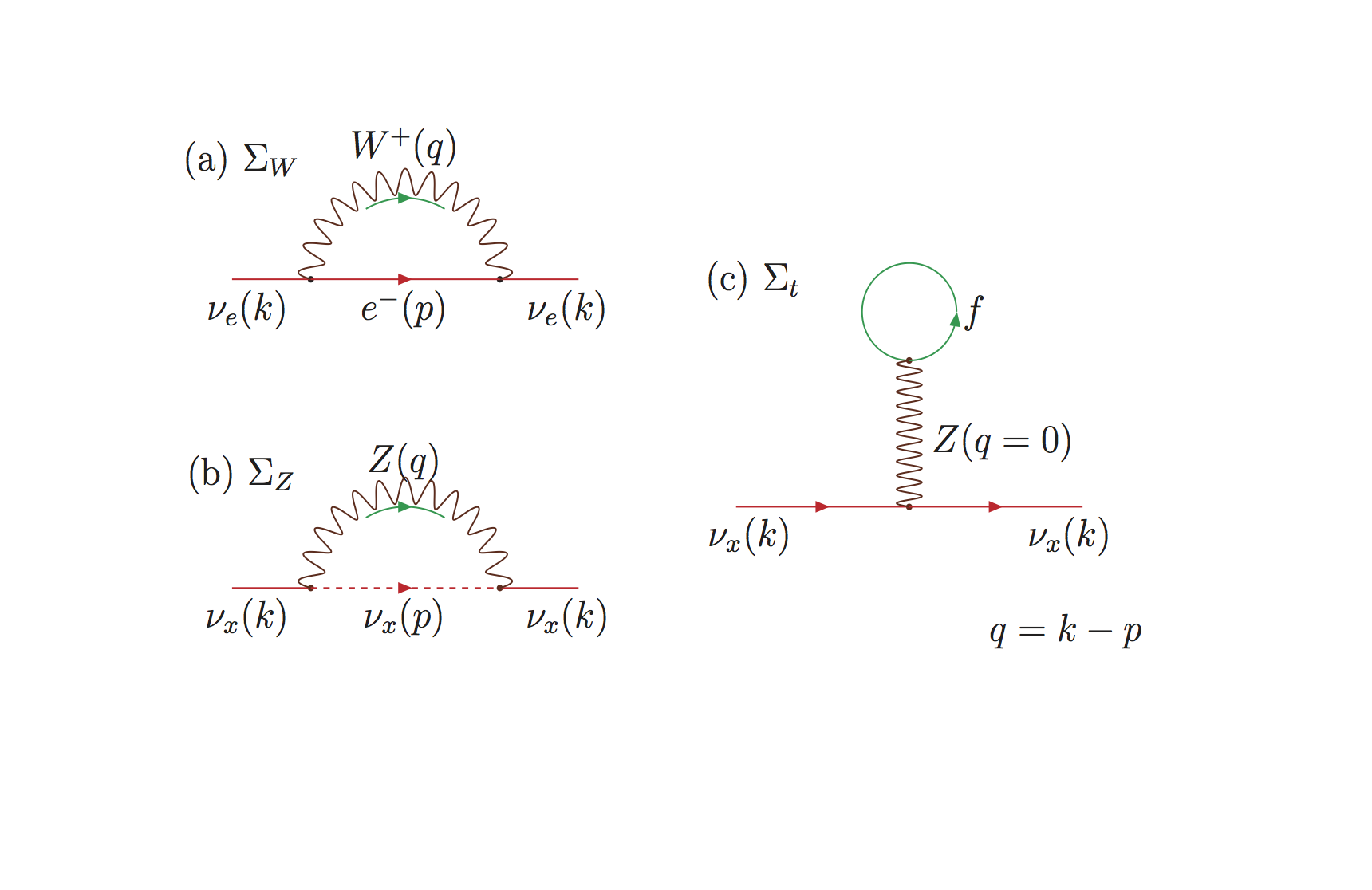}
\caption{One-loop contribution to the neutrino self-energy in a magnetized medium.  a)  W-exchange diagram:  The solid line represents  the electron propagator and the wiggly line the W-boson in a magnetized medium.   b) Z-exchange diagram:  The dashed line corresponds to the neutrino propagator in a thermal medium  and the wiggly line to the Z-boson   c) Tadpole diagram:  The solid line represents  the fermion propagator and the wiggly line the Z-boson in a magnetized medium.}
\label{fig1}
\end{figure*}
\begin{figure*}[htp]
\centering
\includegraphics[width=\textwidth]{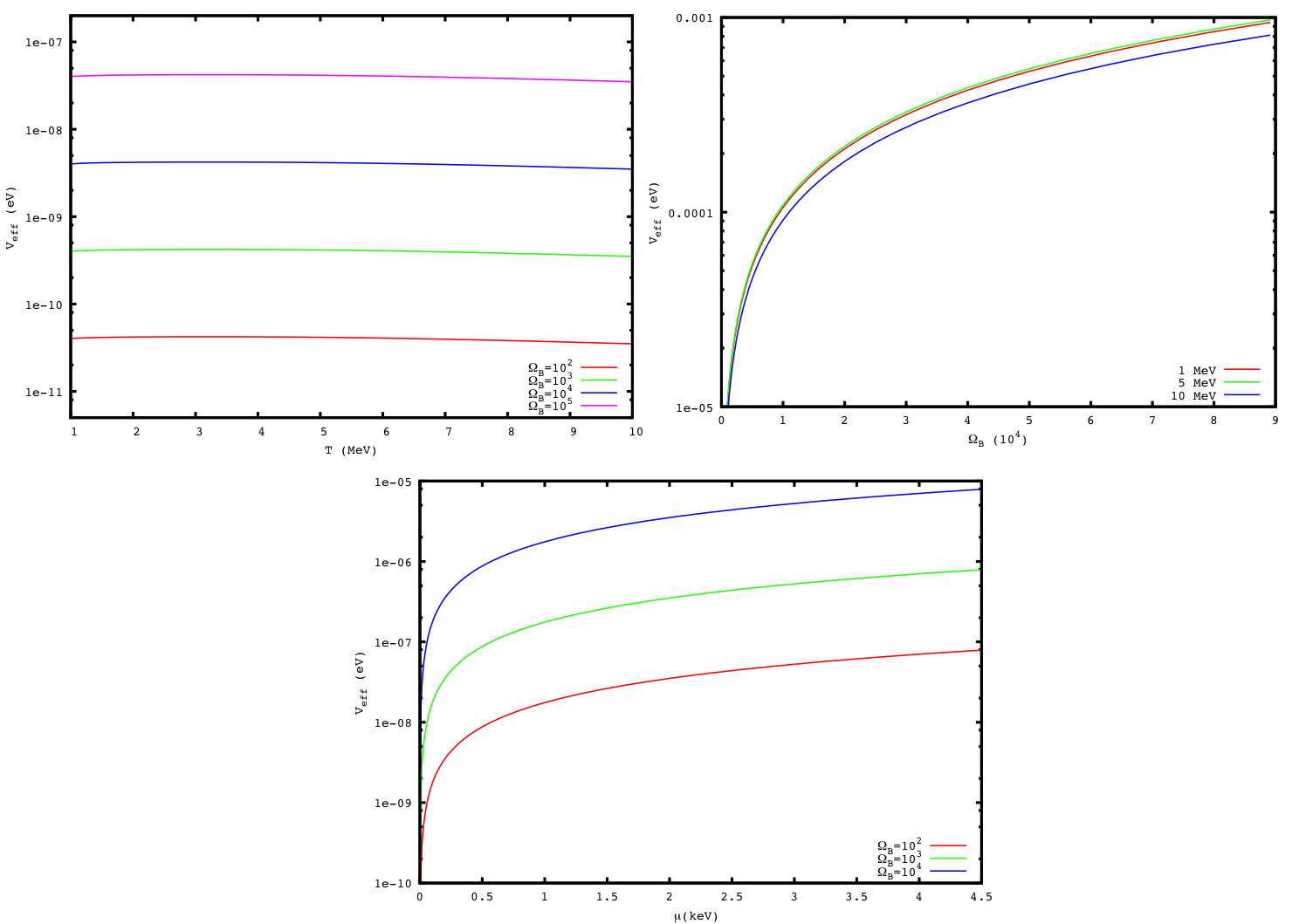}
\caption{Neutrino effective potential in the strong magnetic field regime as a function of temperature (T) (left-hand figure above), magnetic field ($\Omega_B$) (right-hand figure above), chemical potential ($\mu$) (figure below) and neutrino energy (E$_\nu$).  All plots are obtained for a neutrino energy of 10 MeV.} \label{Bstrong}
\end{figure*}
\begin{figure*}[htp]
\centering
\includegraphics[width=\textwidth]{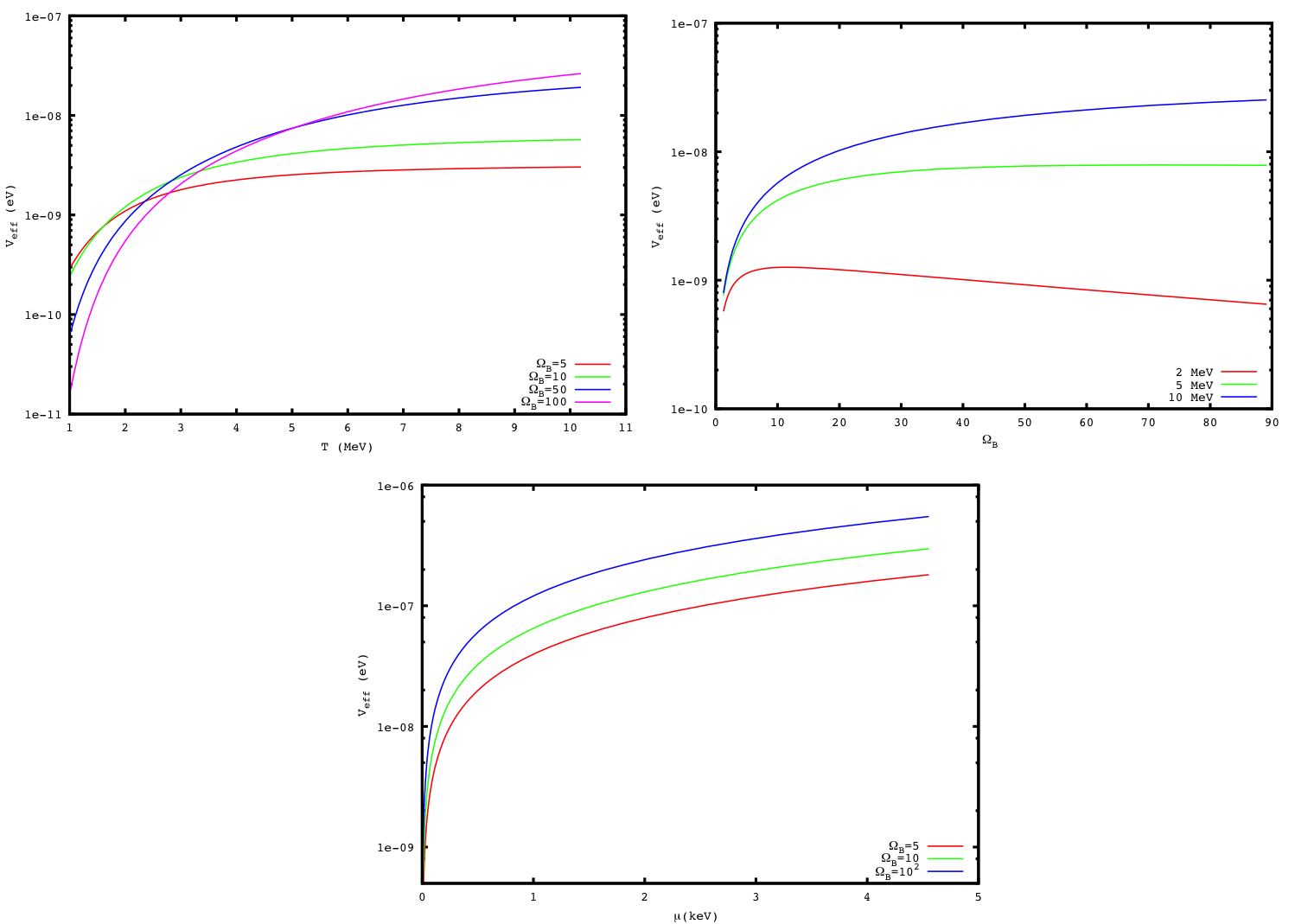}
\caption{Neutrino effective potential in  moderate above  critical magnetic field regime as a function of temperature (T) (left-hand figure above), magnetic field ($\Omega_B$) (right-hand figure above), chemical potential ($\mu$) (figure below) and neutrino energy (E$_\nu$).  All plots are obtained for a neutrino energy of 10 MeV.} \label{Babove}
\end{figure*}
\begin{figure*}[htp]
\centering
\includegraphics[width=\textwidth]{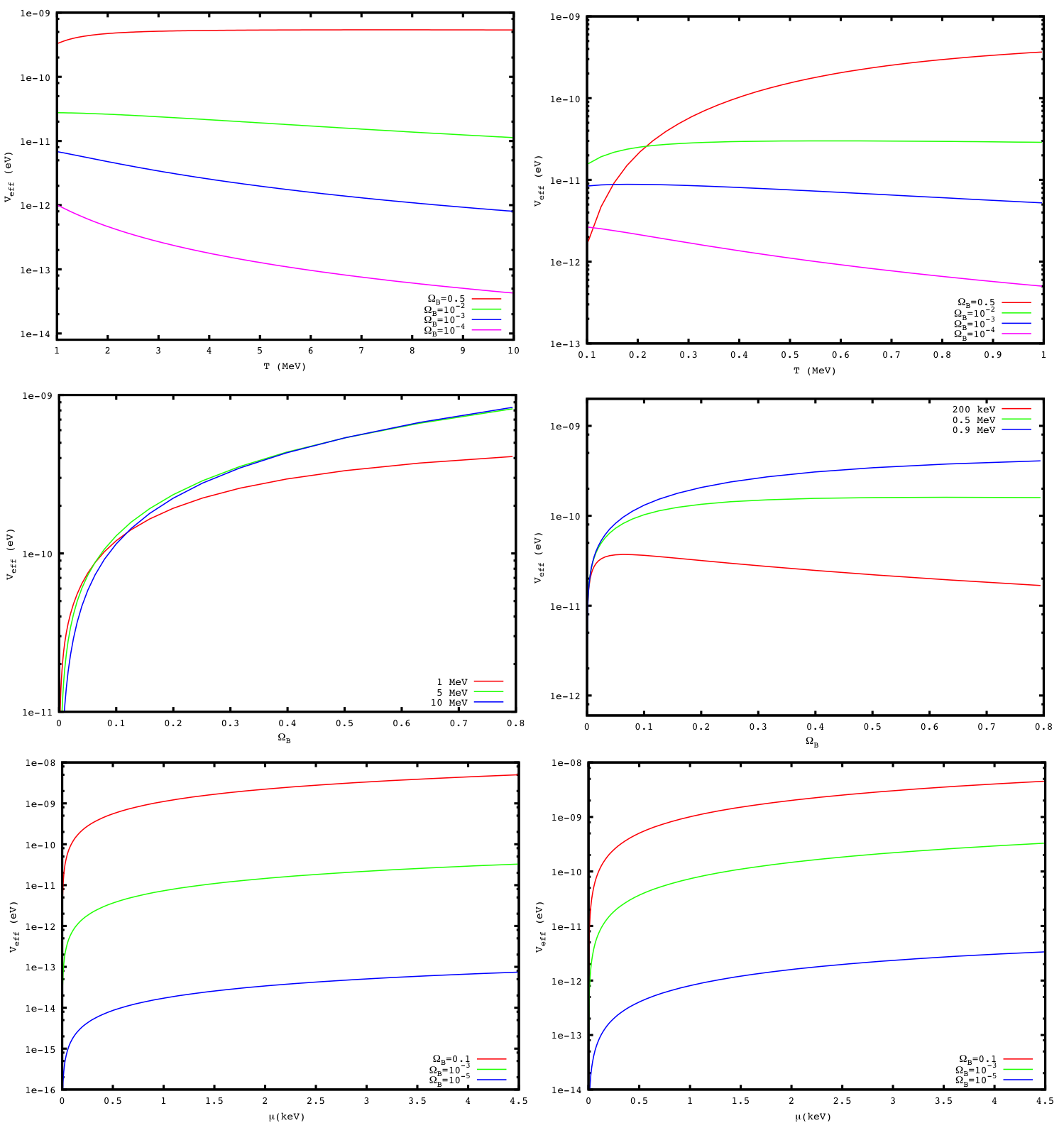}
\caption{Neutrino effective potential in  moderate below  critical magnetic field regime as a function of temperature (T) (top), magnetic field ($\Omega_B$) (middle), chemical potential ($\mu$) (bottom). In the left column a neutrino energy $E_\nu=10$ MeV was used whereas the right column was plotted for  $E_\nu=10$ GeV.}\label{Bbelow}
\end{figure*}
\begin{figure*}[htp]
\centering
\includegraphics[width=\textwidth]{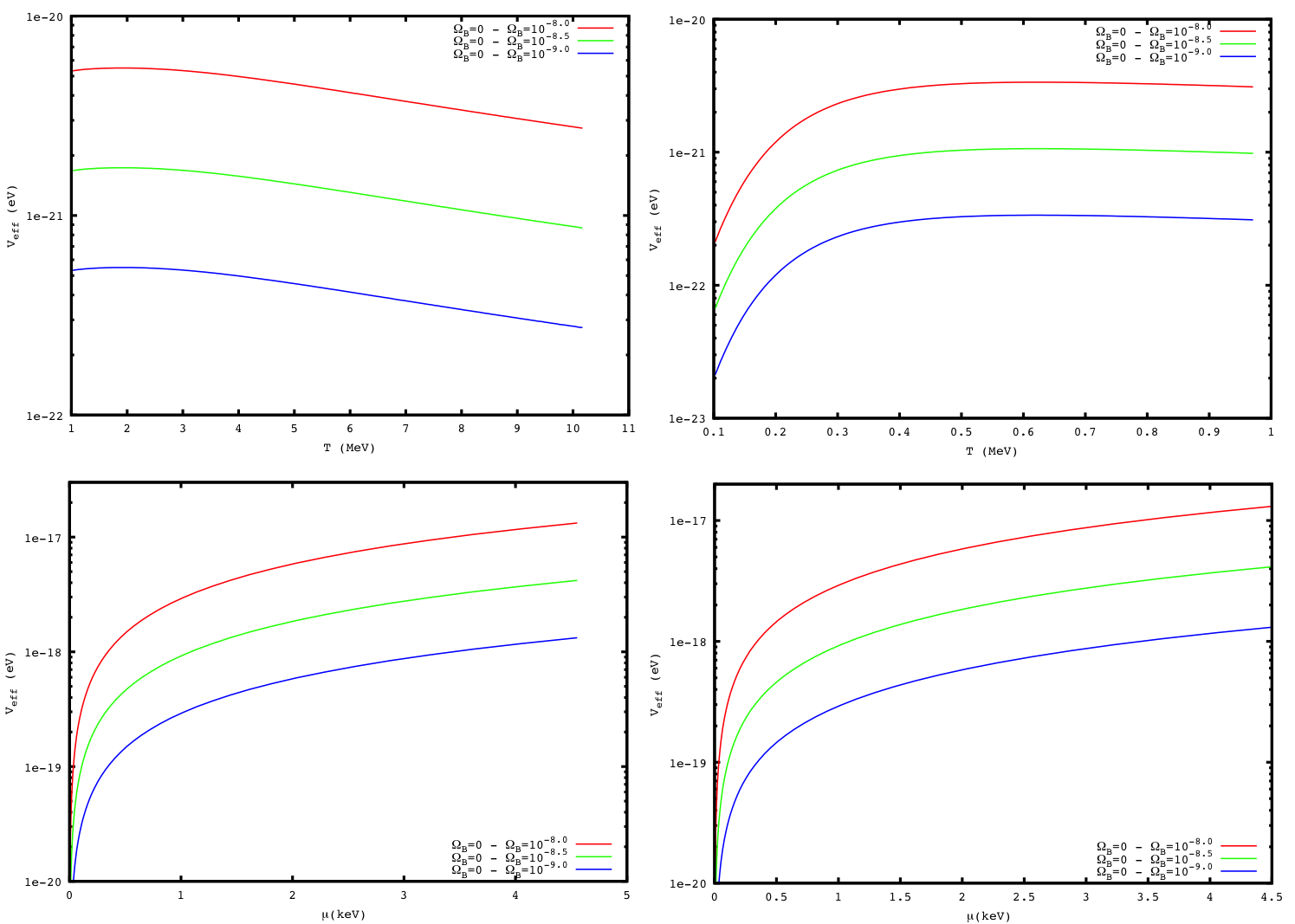}
\caption{Neutrino effective potential in  weak magnetic field regime as a function of temperature (T) (above) and  chemical potential ($\mu$) (below). In the left column a neutrino energy $E_\nu=10$ MeV was used whereas the right column was plotted for  $E_\nu=10$ GeV.} \label{Bweak}
\end{figure*}
\begin{figure*}[htp]
\centering
\includegraphics[width=\textwidth]{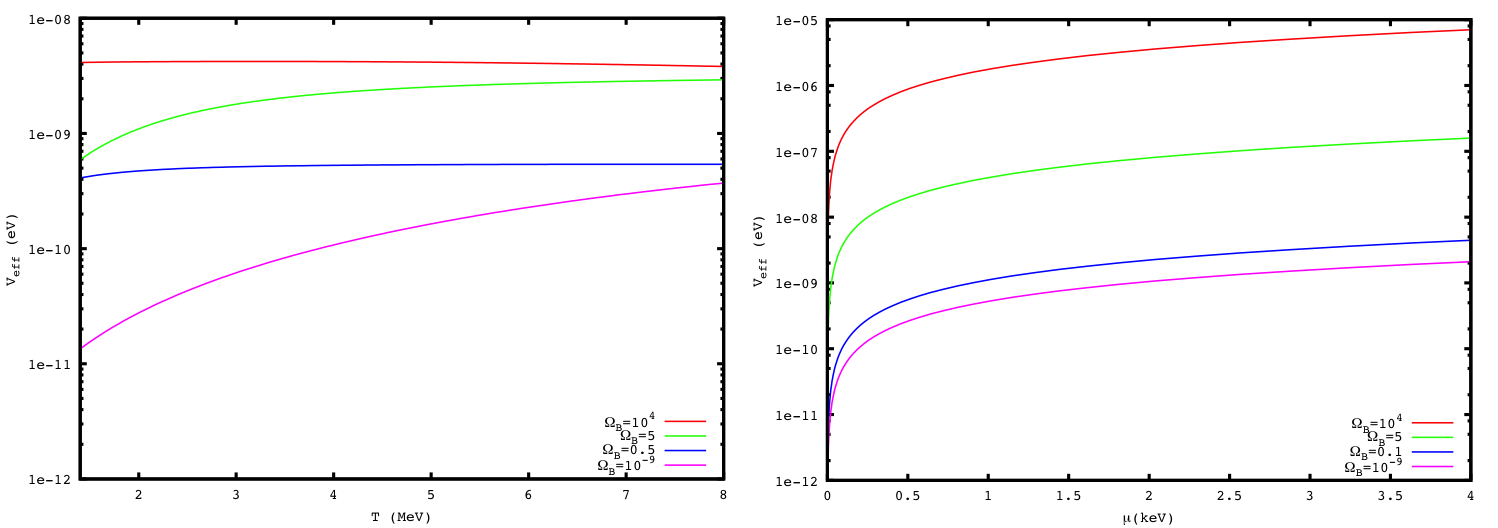}
\caption{Neutrino effective potential as a function of temperature (left) and chemical potential (right) for fixed values of magnetic field in the strong (10$^4 B_c$), moderate (5 and 0.5 $B_c$ for above and below, respectively) and weak (10$^{-9} B_c$) regime.}\label{Bcompar}
\end{figure*}
\begin{figure*}[htp]
\centering
\includegraphics[width=\textwidth]{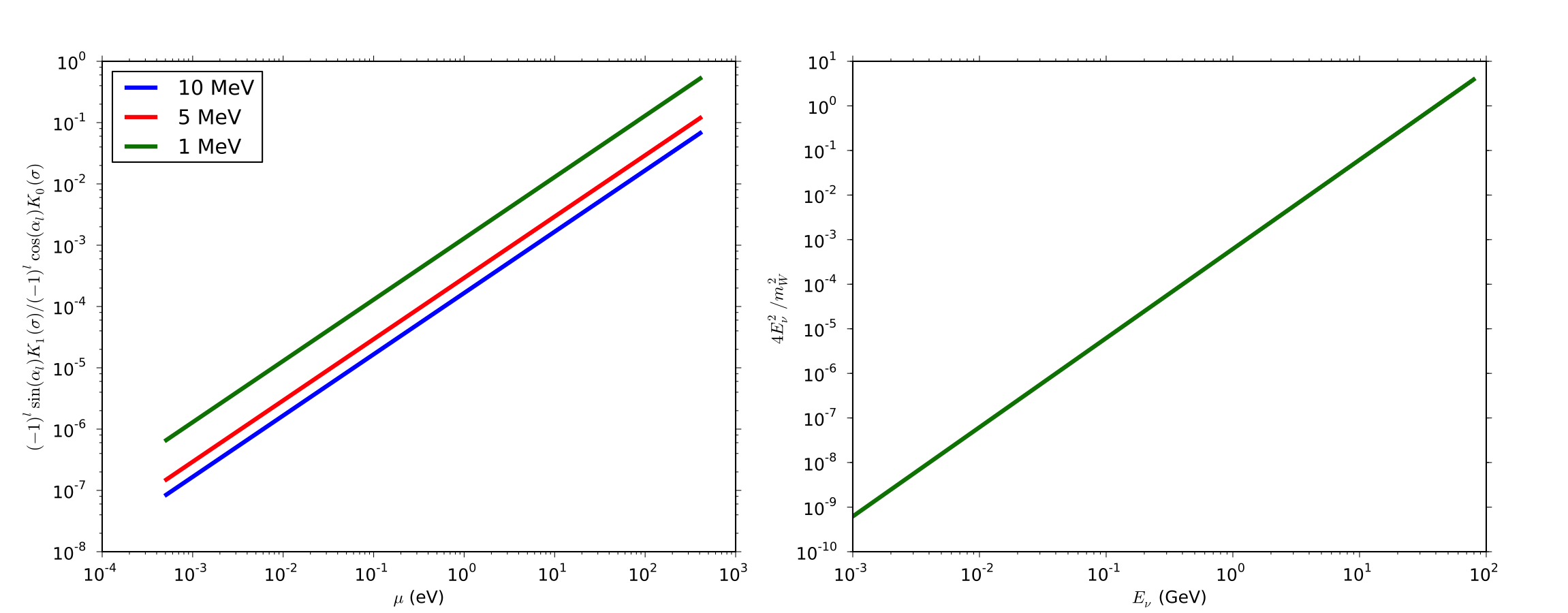}
\caption{Contribution of the terms of order m$^{-4}_W$ in the neutrino effective potential.}
\label{order4}
\end{figure*}
\begin{figure*}[htp]
\centering
\includegraphics[width=\textwidth]{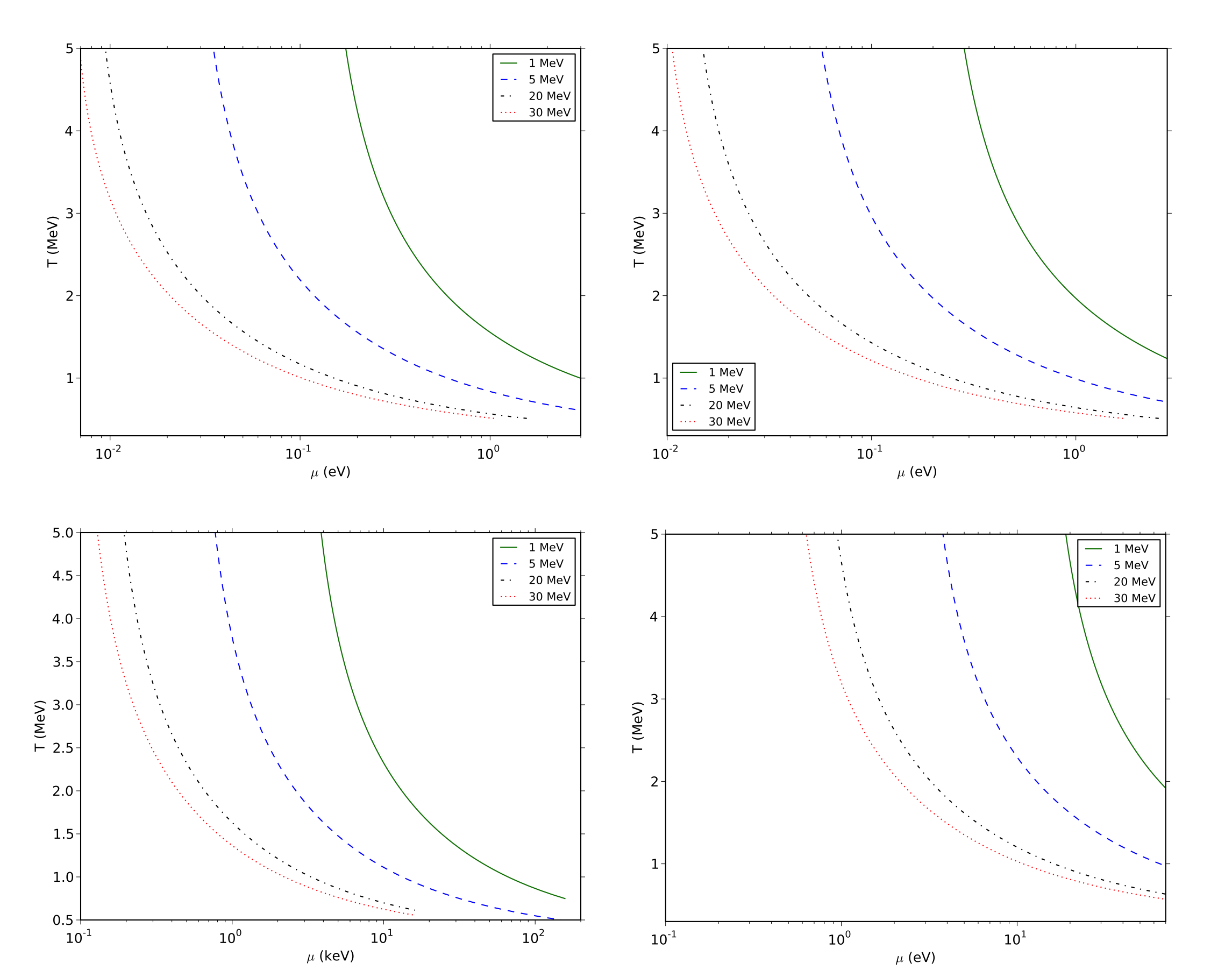}
\caption{Contour plots of temperature  (T) and chemical potential  ($\mu$) as a function of thermal neutrino energy for which the resonance condition is satisfied. We have applied the neutrino effective potential with moderate magnetic field ($\Omega_B =10$) and used the best fit values  of the two; solar (left-hand figure above),   atmospheric (right-hand figure above), and accelerator (left-hand figure below), and three (left-hand figure below) neutrino mixing.}\label{res_ab} 
\end{figure*}
\begin{figure*}[htp]
\centering
\includegraphics[width=\textwidth]{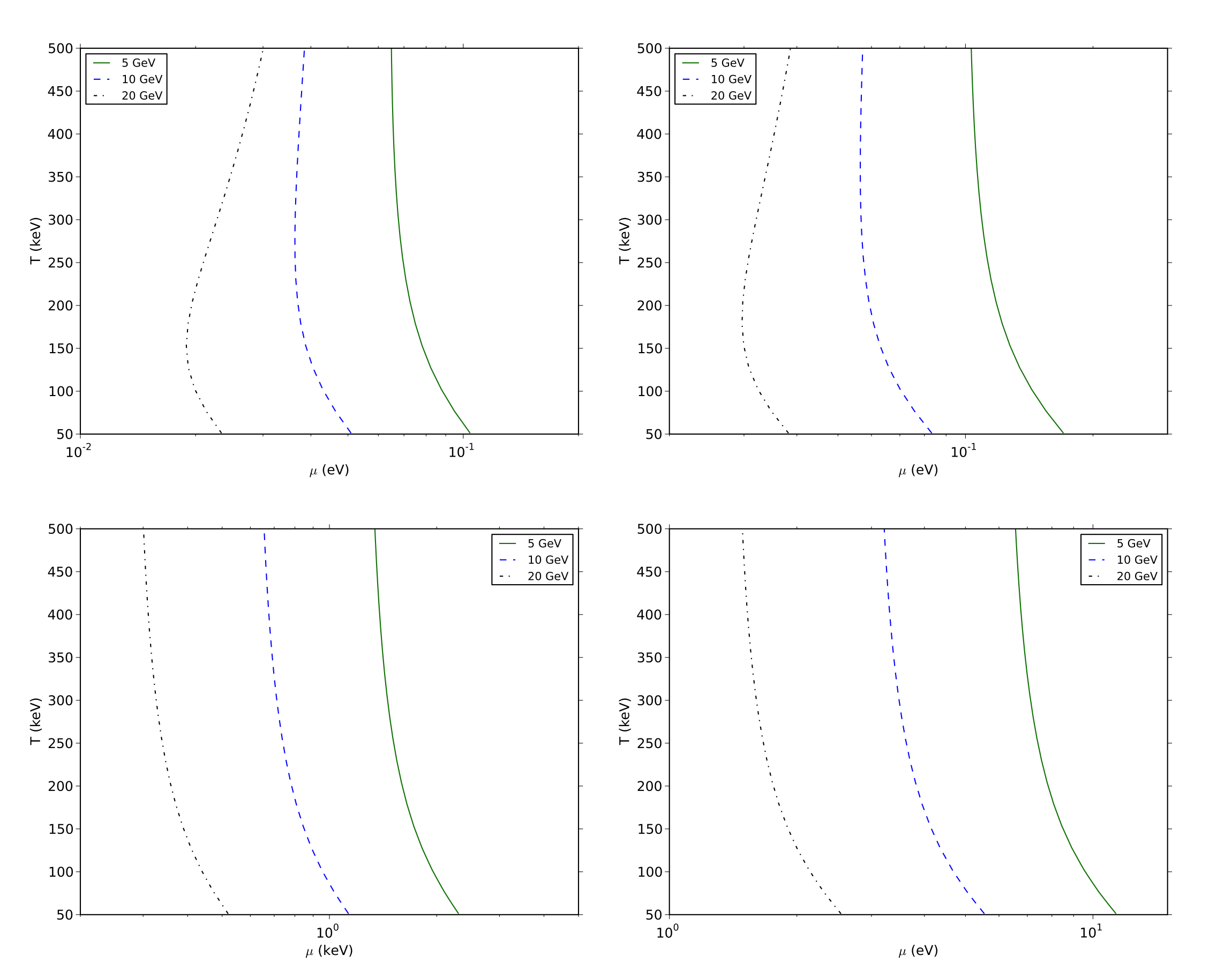}
\caption{Contour plots of temperature  (T) and chemical potential  ($\mu$) as a function of quasi-thermal neutrino energy for which the resonance condition is satisfied. We have applied the neutrino effective potential with moderate magnetic field ($\Omega_B =10^{-4.3}$) and used the best fit values  of the two; solar (left-hand figure above),   atmospheric (right-hand figure above), and accelerator (left-hand figure below), and three (left-hand figure below) neutrino mixing.}\label{res_be}
\end{figure*}
\begin{figure*}[htp]
\centering
\includegraphics[width=\textwidth]{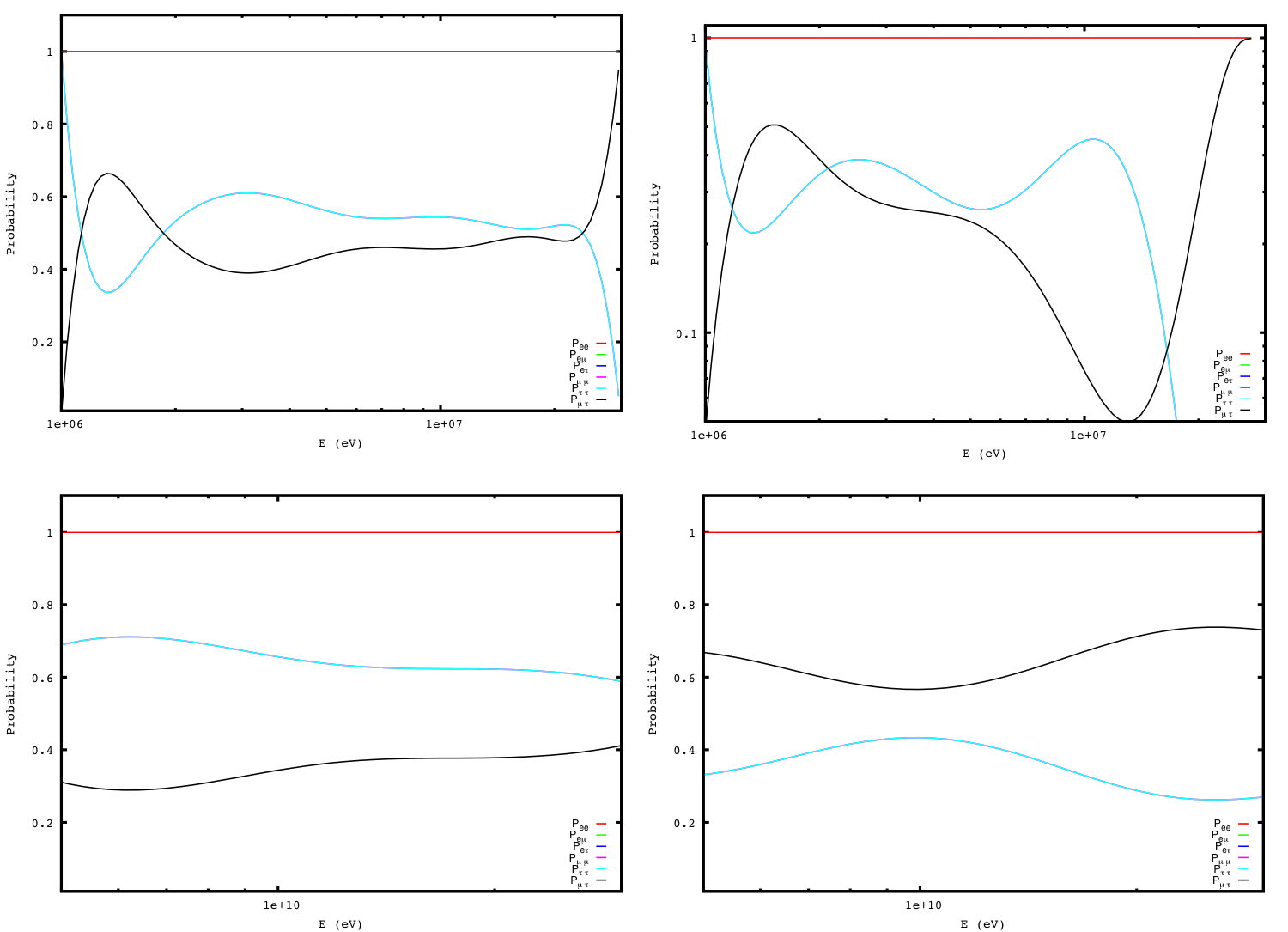}
\caption{Oscillation probabilities of thermal MeV (figures above) and sub-photospheric (figures below) neutrinos  as a function of neutrino energy are plotted. For thermal neutrinos we have used the values of  magnetic field $10\, B_c$ and the fireball radii $10^7$ cm (left figure) and $10^8$ cm (right figure) whereas for sub-photospheric neutrinos we have used the values  B=$10^{-4.3}\,B_c$ and $10^{11}$ cm (left figure) and $10^{12}$ cm, respectively.}
 \label{prob}
\end{figure*}
\begin{figure*}[htp]
\centering
\includegraphics[width=\textwidth]{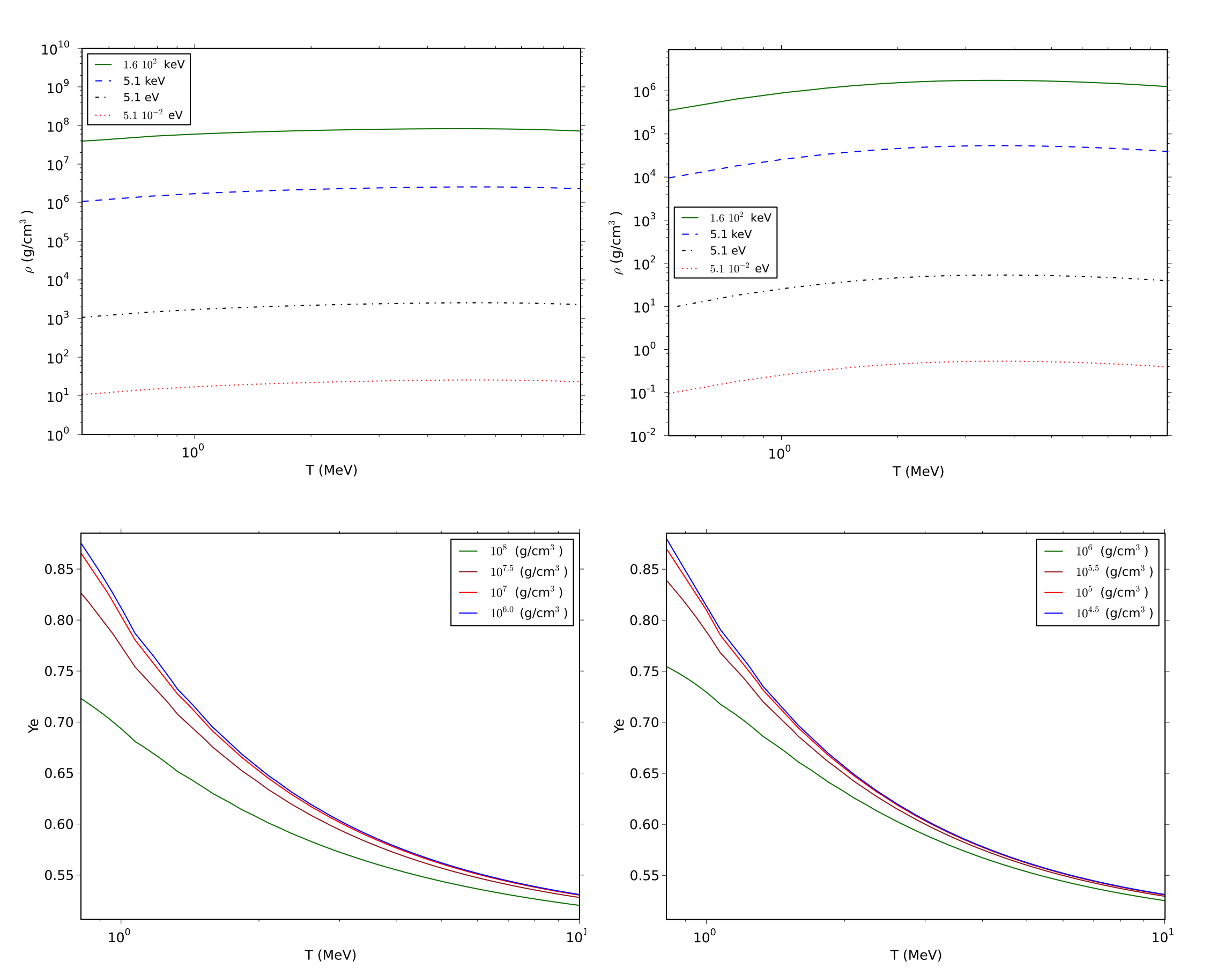}
\caption{Contour plots of baryon density (above) and  Ye  (below)   as a function of  temperature    for which the equilibrium condition is established. We use the values of  temperature and chemical potential in the resonance condition range and moderate field limit:  50 $B_c$ (left-hand figures) and  0.1 $B_c$ (right-hand figures).}
\label{ye} 
\end{figure*}
\begin{figure*}[htp]
\centering
\includegraphics[width=\textwidth]{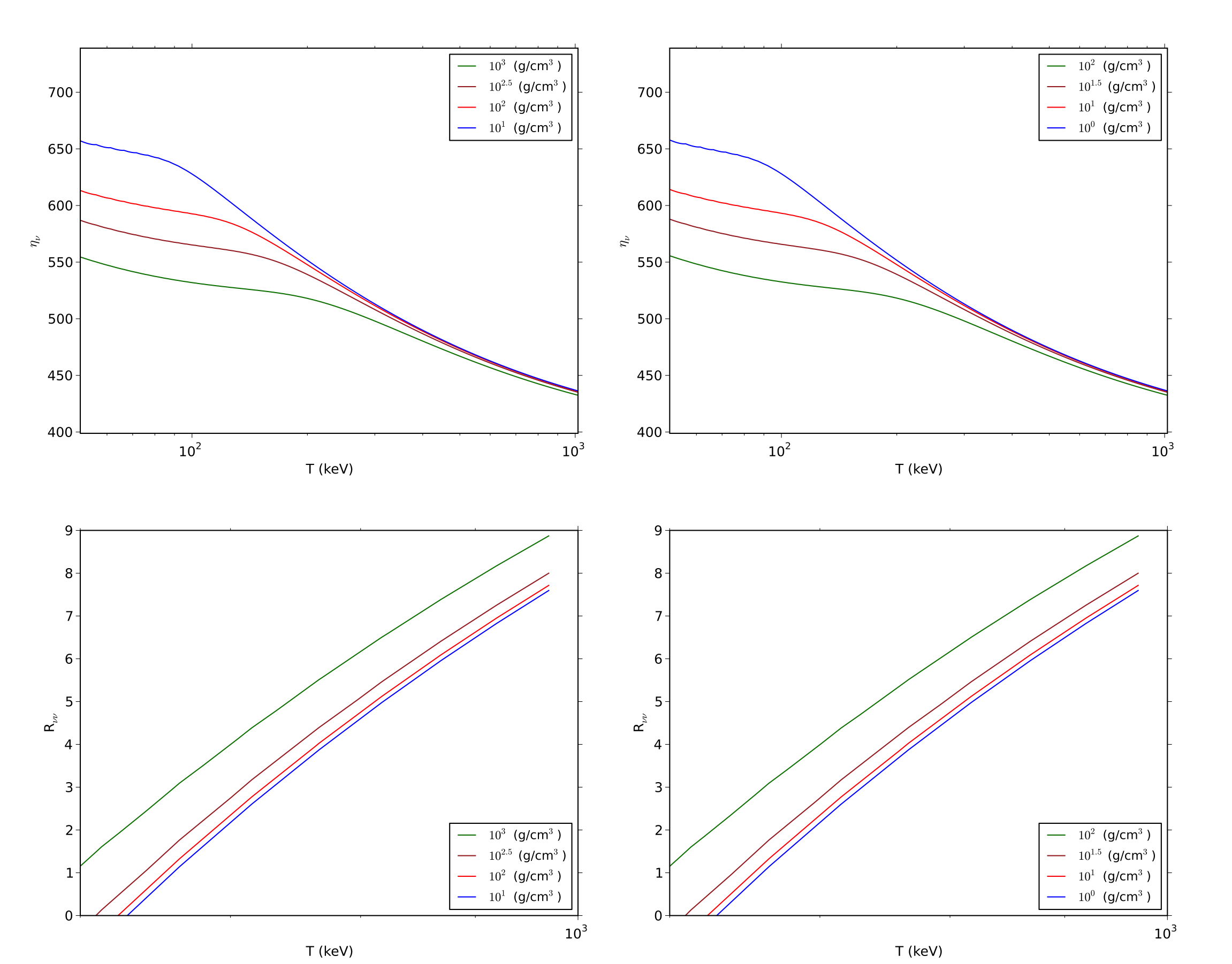}
\caption{Contour plots of  the critical value  of baryon load parameter($\eta_\nu$) (above) and  number of expected events ($R_{\nu\nu}$) (below) as a  function of and temperature (T) \citep{mes00}.  We use the values of  temperature and chemical potential in the resonance condition range and moderate field limit:   $10^{-4.3}\, B_c$ (left-hand figures) and  $10^{-5.3}\,B_c$ (right-hand figures). The number of events is calculated for z=0.1}
\label{neut}
\end{figure*}
\end{document}